\documentclass[useAMS,usenatbib]{mn2e}
\usepackage{graphicx,amssym}

\newcommand{\bc}{\begin{center}}
\newcommand{\ec}{\end{center}}

\title[Cosmic FIR/sub-mm background]
      {Contribution of the first galaxies to the cosmic far-infrared/sub-millimeter background -- I.
       Mean background level}
\author[M. E. De Rossi and V. Bromm]
       {\parbox{17cm}{Mar\'ia~Emilia De Rossi$^{1,2}$\thanks{Email:mariaemilia.dr@gmail.com} and Volker Bromm$^{3}$}
       \\     
       \\
       $^{1}$ Consejo Nacional de Investigaciones Cient\'ificas y T\'ecnicas, Argentina\\  
       $^{2}$ Instituto de Astronom\'{\i}a y F\'{\i}sica del Espacio (IAFE, CONICET-UBA), CC 67, Suc. 28, 1428 Buenos Aires, Argentina \\
       $^{3}$ Department of Astronomy, University of Texas at Austin, 2511 Speedway, Austin, TX 78712, USA 
}
\begin{document}

\date{Accepted  ???? ??. 2016 ???? ??}

\pagerange{\pageref{firstpage}--\pageref{lastpage}} 
\pubyear{2016}

\maketitle

\label{firstpage}

\begin{abstract}
We study the contribution of the first galaxies to the far\--infrared/sub\--millimeter (FIR/sub\--mm) extragalactic 
background light (EBL) by implementing an analytical model for dust emission. 
We explore different dust models, assuming different grain size distributions and chemical compositions.
According to our findings, observed re-radiated emission from dust in dwarf-size galaxies
at $z \sim 10$ would peak at a wavelength of $\sim 500 \mu {\rm m}$ with
observed fluxes of $\sim 10^{-3} \-- 10^{-2}$ nJy, which
is below the capabilities of current observatories.
In order to be detectable, model sources at these high redshifts should exhibit
luminosities of $\gtrsim 10^{12} L_{\sun}$, comparable to that of local ultra-luminous systems.
The FIR/sub\--mm EBL generated by primeval galaxies
peaks at $\sim 500 \mu {\rm m}$, with an intensity ranging from
$\sim 10^{-4}$ to $10^{-3} {\rm nW \ m^{-2} \ sr^{-1}}$, depending on dust properties. These values are
$\sim 3 \-- 4$ orders of magnitude below the absolute measured cosmic background 
level, suggesting that the first galaxies would not contribute significantly to the observed FIR/sub\--mm EBL.
Our model EBL exhibits a strong correlation with 
the dust-to-metal ratio, where we assume a fiducial value of $D = 0.005$, 
increasing almost proportionally to it.
Thus, measurements of the FIR/sub\--mm EBL could provide constraints
on the amount of dust in the early Universe. Even if the absolute
signal from primeval dust emission may be undetectable, it might still
be possible to obtain information about it by exploring angular fluctuations
at $\sim 500 \mu {\rm m}$, close to the peak of dust emission from the first galaxies.
\end{abstract}

\begin{keywords}                                                                                                     
cosmology: theory -- galaxies: evolution -- galaxies: abundances -- 
galaxies: haloes -- galaxies: high-redshift -- 
galaxies: star formation
\end{keywords}

\section{Introduction}
\label{sec:introduction}

The emergence of the first stellar systems fundamentally transformed the Universe 
from its simple initial state during the cosmic dark ages into one of progressively larger complexity 
\citep[e.g.][]{barkana2001, loeb2010}.
Besides contributing to the reionization of the Universe, primordial stars 
enriched the pristine gas with the first heavy chemical elements \citep[e.g.][]{bromm2003, 
furlanetto2003, yoshida2004}, including the formation
of dust \citep{schneider2006_methods, 
cherchneff2010, chiaki2015}.
The complex physics of pre-galactic metal enrichment \citep[e.g.][]{karlsson2013},
and the properties of primordial dust inside the first cosmic structures
\citep[e.g.][]{loeb1997, schneider2004, gall2011} constitute
frontier topics in modern cosmology as they are concerned with the foundation
for subsequent galaxy formation.

Within the standard $\Lambda$-Cold Dark Matter ($\Lambda$CDM) model, the
first stellar populations are predicted to have formed from metal-free gas 
inside dark matter minihaloes ($\sim 10^6 M_{\sun}$) at $z \gtrsim 20$ 
\citep[e.g.][]{couchman1986, haiman1996, tegmark1997, yoshida2003, 
bromm2004, bromm2009, bromm2013}.
The first stars, the so-called Population~III (Pop~III), exhibited high characteristic masses 
($\sim 10-100 M_{\sun}$), and had correspondingly short lifetimes \citep[e.g.][]{abel2002, bromm2002, 
stacy2010, greif2011}. Pop~III stars, therefore, may have been important sources of dust at early epochs.  
Owing to the shallow gravitational potential wells of minihaloes,
the strong negative feedback from Pop~III stars probably heated and expelled 
all remaining gas, inhibiting further star formation (SF).
This implies that Pop III minihaloes would not have been true galaxies, 
if they are defined as the hosts of long-lived stellar systems. 
The deeper potential wells of atomic cooling haloes, with virial masses of $\sim 10^7 \-- 10^8 M_{\sun}$,
are promising candidates for hosting second-generation metal-poor (Pop~II) stars
at $z \sim 20 \-- 6$ \citep[][]{bromm2011, bromm2011b}. Hence, these systems
would be the first protogalaxies in the pre-reionization Universe.
At this epoch, the SF process transitioned into a low-mass dominated mode, described by a more normal initial mass function (IMF), similar to the near-universal
Salpeter law \citep[e.g.][]{frebel2007, ji2014}. This transition was likely
driven by a combination of
atomic fine structure cooling and dust thermal emission. Thus, the
first galaxies might have been composed of Pop~II stellar systems,
surrounded by a mixed phase of gas and dust inside atomic cooling haloes.

In modern cosmology, a challenging issue is to understand the interconnection 
between the diffuse Extragalactic Background Light (EBL) and the first luminous systems, 
specifically to determine to what extent this radiation could be attributed to sources in the early Universe.
Primordial stellar populations are expected to have left a measurable
imprint on the absolute optical and near-infrared (NIR, $\lambda \sim 1 \-- 10 \mu {\rm m}$) EBL \citep[e.g.][]{santos2002}, 
and its spatial fluctuations \citep[e.g.][]{cooray2004, kashlinsky2004}.
As photons propagate towards the observer in an expanding Universe, they lose
energy and any radiation associated with the first stellar systems will be seen mostly
in the NIR \citep[][]{bromm2013}, peaking at about $1 \mu {\rm m}$. 
In particular, HII regions enclosing ionizing sources in the first luminous systems 
contribute to the cosmic near-infrared background
by redshifted Lyman-$\alpha$ emission \citep[e.g.][]{salvaterra2003}.
On the other hand, UV radiation from primordial stars that heated interstellar dust and
was re-radiated at longer wavelengths contributed to the EBL
in the far-IR/sub-millimeter (FIR/sub-mm)
part of the spectrum \citep[e.g.][]{low1968, kaufman1976, carr1984, beichman1991, dwek1998}.

The origin of the NIR-EBL, often succinctly referred to as the ``cosmic infrared background'' (CIB), has been widely discussed in the literature.
Specifically, the measured excess in the NIR-EBL over what is expected from known galaxies 
might be indicating a significant contribution from early epochs \citep[e.g.][]{kashlinsky2005}.
However, because of the difficulties to measure absolute flux due to the strong foregrounds,
CIB studies have mainly focused on the determination of spatial fluctuations.
Recently, \citet{mitchellwynne2015} performed a multiwavelength study
by using one of the deepest surveys with the Hubble Space Telescope. In analysing
CIB spatial fluctuation at ${\lambda} = 0.6 \-- 1.6 \mu {\rm m}$, they
infer a significant surface density of faint sources at $z>8$, which are still below
the point-source sensitivities of current observatories.
This result is encouragingly consistent to previous findings deduced by \citet{kashlinsky2007} 
by using {\it Spitzer} data.
With the advent of new instruments such as the NIRCam on the {\it James Webb Space Telescope (JWST)},
the detection of individual, fainter systems will become possible, resolving
the sources of the CIB to
even higher $z$.

With respect to the FIR/sub-mm EBL,
observational studies benefit from the negative K-correction. 
Sources associated with this wavelength range are generally dusty galaxies,
where surrounding dust absorbs optical/UV emission from young stars, to be re-radiated in the FIR.
During the last two decades, the SCUBA and LABOCA surveys resolved 20\--40\% of the EBL at 
850 $\mu {\rm m}$ \citep[e.g.][]{eales1999, Coppin2006, weiss2009}, and 10\--20\% at 
$1$ mm in deep surveys with the AzTEC camera \citep{Wilson2008}. 
In recent years, it was suggested that bright dusty high-$z$ sub-mm galaxies \citep[SMGs;][]{casey2014}
can account for more than half of the EBL at $850 \mu {\rm m}$ \citep{viero2013, cai2013}.
However,
SMGs constitute an extreme case and cannot be representative of the bulk of the galaxy
population at high $z$. 
The major contribution to the observed flux at mm/sub-mm wavelengths seems to originate in fainter  
($< 1$mJy), high-$z$ populations different from SMGs.
Very recently, the higher sensitivity of the Atacama Large Millimeter/submillimeter Array (ALMA) 
allows to infer the number counts at fluxes fainter than 1 mJy.
\citet{hatsukade2013} resolved 80\% of the EBL at $1.3$ mm, 
when exploring faint (0.1\--1 mJy) objects. Similar findings were reported by \citet{ono2014} 
at $1.2$ mm. 
In addition, \citet{carniani2015} identified 50 sources down to 60 $\mu$Jy,
claiming that the observed flattening of the integrated number counts at faint
fluxes might be indicating that they are close to resolving 100\% of the cosmic background, subject to 
large uncertainties regarding the absolute EBL level.

Although the contribution of the first stellar systems to the NIR-EBL has been extensively
studied \citep[e.g.][]{santos2002, salvaterra2003, salvaterra2006, kashlinsky2002,kashlinsky2004,
kashlinsky2005b, magliocchetti2003, cooray2004, cooray2004b}, their role
as possible FIR/sub-mm sources is still poorly explored. In this paper, 
we ask to what extent the first galaxies may have contributed
to the observed FIR/sub-mm EBL through redshifted dust re-emission. To accomplish this,
we implement an idealized analytical model
designed to reproduce properties of the first galactic systems.
This  model is then combined with a Sheth-Tormen halo mass function to estimate
their contribution to the EBL at FIR/sub-mm wavelengths.

In Section~2, we discuss our primeval galaxy models in detail, to be
followed in Section~3 and 4 with the description of our background light
modelling. We conclude in Section~5 with a brief outlook into future
developments.
The following cosmological parameters are assumed in this work:
$h$ = 0.67, 
${\Omega}_{\rm b}$ = 0.049,
${\Omega}_{\rm M}$ = 0.32,
${\Omega}_{\Lambda}$ = 0.68
\citep{planck2014}.

\section{Dust Emission from Primeval Galaxies}
\label{sec:dust_model}

We developed an idealized model to explore dust emission signatures
associated with haloes hosting the first Pop~II stellar populations.
Observed specific fluxes were calculated for individual sources
of different masses and at different redshifts. 
In Section~3, we will
combine these models
with the Sheth-Tormen halo mass function
to predict the observed cosmic background in the FIR/sub-mm range.
We now proceed to describe the model and our choice of input parameters.

\subsection{Model galaxies}

We consider the idealized case of a dark-matter halo hosting
a central stellar cluster, surrounded by a mixed phase of gas
and dust that extends towards the virial radius ($R_{\rm vir}$). 
We assume spherical symmetry inside these systems, and do not
consider any extended distribution of halo stars.
To construct the density profile for the {\em gaseous} component, 
we adopt an isothermal power law of the form $\sim r^{-2}$, 
as it provides a good description of a virialized system \citep{binney2008}. 
Following \citet{smith2015}, we
consider a non-cuspy 'core', consistent with observations of 
low surface brightness galaxies \citep[e.g.][]{kormendy2009}.
Thus, the gas mass density profile has the form:
\begin{equation}\label{eq:rhog_vs_r}
{\rho}_{\rm g} (r) \propto \left\{ 
\begin{array}{ll}
1       &      r \le R_{\rm c}  \\
{(\frac{R_{\rm c}}{r})}^2      &       R_{\rm c} < r \le R_{\rm vir} 
\end{array} \right.
\end{equation}
where $R_{\rm c}$ denotes a 'core' radius within which the density flattens
off to a constant value. 
By construction, this density profile satisfies the following condition:
\begin{equation}
1<F_{\rm g} = \frac{M_{\rm vir, g}}{ \frac{4}{3} \pi  R_{\rm vir}^3  {\rho}_{\rm g}(r = R_{\rm vir}) }<3.
\end{equation} 
where $M_{\rm vir, g}$ is the total gas mass contained inside $R_{\rm vir}$.

For a given halo virial mass ($M_{\rm vir}$) and redshift ($z$),
$R_{\rm vir}$ is defined as the radius which encloses an overdensity of
200 with respect to the cosmic mean, $\bar{\rho} (z)$, at that redshift
\citep{bromm2011b}.
The parameter $R_{\rm c}$ and the central gas density are set by imposing
two additional conditions. Firstly, for  
the baryon-to-total mass ratio of the system to be of order
the cosmic mean ${\Omega}_{\rm b} / {\Omega}_{\rm M}$, 
we require that ${\rho}_{\rm g} (R_{\rm vir}) = 200  
({{\Omega}_{\rm b}} / {{\Omega}_{\rm M}}) \bar{\rho} (z)$. 
We are assuming 
that the stellar contribution is negligible, which should be valid for the first galaxies.
Secondly, we fix $F_{\rm g}$ in such a way that central
gas densities are consistent with the expectation for the first galaxies
($\gtrsim 1 {\rm \,cm}^{-3}$), resulting in $F_{\rm g} \approx 2.8$.
By construction, in this model, 
central densities are similar at similar redshifts, regardless of halo mass.
For our selected parameters, they range from $n_{\rm g} \sim 2.0 \ {\rm cm}^{-3}$ at $z=7$ to $n_{\rm g} \sim 40 \ {\rm cm}^{-3}$
at $z=20$.
\begin{figure}
\begin{center}
\resizebox{7.5cm}{!}{\includegraphics{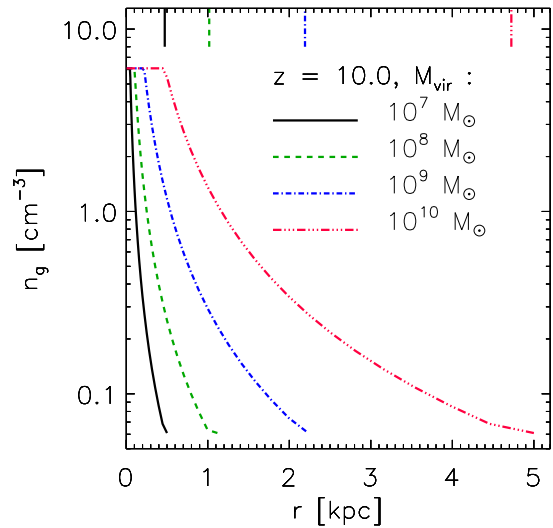}}
\end{center}
\caption[ng vs r at $z=10$]
{
Gas density profiles for model galaxies with different $M_{\rm vir}$ at $z=10$.
The vertical lines on the top depict the virial
radii of these systems.
Note that the vertical axis employs logarithmic scaling.
}
\label{fig:ng_vs_r_z10}
\end{figure}
In Fig. \ref{fig:ng_vs_r_z10}, we show sample density profiles for systems
of different mass at $z=10$.
We derive the dust density distribution inside haloes
from the gas density by assuming a dust-to-metal
mass ratio $D = M_{\rm d} / M_{\rm Z} = 5 \times 10^{-3}$,
and a gas metallicity of 
$Z_{\rm g} = 5 \times 10^{-3} Z_{\odot}$.
\footnote{We adopt $Z_{\odot}=0.0127$ \citep{allendeprieto2001}, which
corresponds to the default in {\sc CLOUDY} \citep[version 07.02,][]{ferland1998}.
This value is also consistent with that reported by \citet[][$Z_{\odot}=0.0122$]{asplund2005} and,
more recently, by \citet[][$Z_{\odot}=0.0134$]{asplund2009}.}

As our model galaxies include a primordial dust component, their stellar populations
should have already been enriched with metals.
For simplicity, we assume that a compact cluster of Pop~II stars resides
in the center of the host halo, modelled subsequently as a point source
of stellar radiation.
The cluster is assumed to have been formed during a single instantaneous burst
following a Kroupa (2001) IMF, which is a good approximation for Pop~II stars
inside the first galaxies \citep{safranek2010, safranek2014}.
In order to estimate its total stellar mass ($M_*$), we assign a star formation efficiency of 
$\eta = M_* / (M_{\rm g} + M_*) = 0.01$ to the system \citep{greif2006, mitchellwynne2015}.
We construct the
spectral energy distribution (SED) associated with this system
using Yggdrasil model grids \citep{zackrisson2011}. Specifically, we
adopt models corresponding to their lowest available stellar metallicity,
$Z_* \approx 3 \times 10^{-2} Z_{\odot}$, and a stellar age $\tau = 0.01 {\rm \,Myr}$.
In this case, Yggdrasil employs SEDs derived from Starburst99 single stellar population (SSP) models, based on Padova asymptotic giant branch (AGB)
tracks \citep{leitherer1999, vazquez2005}, with a Kroupa IMF in 
the interval 0.1-100 $M_{\odot}$.

\subsection{Dust model}

In modelling the dust chemical composition and grain size
distribution, we follow \citet{ji2014}, who considered a suite of
different silicon-based dust models \citep[see table~1 in][]{ji2014},
calculated by \citet{cherchneff2010}: UM-ND-20, UM-ND-170,
UM-D-20, UM-D-170, M-ND-20, M-ND-170, M-D-20 and M-D-170.
\footnote{Model names refer to the type of dust model from \citet{cherchneff2010}:
UM = unmixed, M = mixed; ND = non depleted, D = depleted; 170: 170 $M_{\sun}$ progenitor,
20: 20 $M_{\sun}$ progenitor.}  
As in \citet{ji2014}, we will not try to address which prescription
is more realistic and take all of them as plausible variations
in the chemical composition of dust.  
Representative species in these models are ${\rm SiO}_2$ , ${\rm Mg}_2 {\rm SiO}_4$, 
amorphous Si, and FeS.
It is worth mentioning that \citet{cherchneff2010} estimated dust chemistry considering non-equilibrium 
chemical kinetics for dust formation; instead, most steady-state models would predict a dominant carbon dust 
composition.
In addition, in the context of the \citet{cherchneff2010} models, 
the suppression of carbon dust relies on the hypothesis
that carbon-rich regions in the supernova ejecta are microscopically mixed with helium ions. 
We here follow these assumptions, but we acknowledge that there is an
ongoing debate about
the extent to which carbon dust formation is 
inhibited \citep[e.g.][]{nozawa2013}. However, our overall conclusions would
remain valid also for a dust composition that contains a carbon-based component
as well.

For the grain size distribution, we consider the simple, well-motivated
``standard'' prescription of \citet{pollack1994}. This was used
in \citet{omukai2010}, and it is similar to the Milky
Way grain size distribution used in, e.g., \citet{dopcke2013}. 
For spherical dust grains of radius $a$, the distribution is:
\begin{equation} \label{eq:stdsizedistr}
  \frac{dn_{\rm standard}}{da} \propto \left\{ \begin{array}{ll}
    1 & a < 0.005 \mu {\rm m} \\
    a^{-3.5} & 0.005 \mu {\rm m} < a < 1 \mu {\rm m} \\
    a^{-5.5} & 1 \mu {\rm m} < a < 5 \mu {\rm m}
  \end{array} \right. \mbox{\, .}
\end{equation}
We also explored the shock size distribution used by \citet{ji2014}:
\begin{equation} \label{eq:shocksizedistr}
  \frac{dn_{\rm shock}}{da} \propto \left\{ \begin{array}{ll}
    1 & a < 0.005 \mu {\rm m} \\
    a^{-5.5} & a> 0.005 \mu {\rm m} 
  \end{array} \right. \mbox{\, .}
\end{equation}

The latter distribution approximates the effect of running a post-supernova reverse
shock through newly created dust and is based on the work of \citet{bianchi2007}.
For simplicity, each type of dust grain is assumed to have the same
grain size distribution.  These models for dust chemistry and grain sizes
are then used to obtain dust geometrical cross sections and dust opacities
as described in \citet[][]{ji2014}, and we refer the reader to that paper for details.

\subsubsection{Dust temperature profile}
\label{sec:td}

To determine the dust temperature ($T_{\rm d}$), we assume thermal
equilibrium. We set the dust cooling rate, ${\Lambda}_{\rm d} (T_{\rm d})$,
equal to the dust heating rate driven by gas-dust
collisions, $H_{\rm d}$, and by the stellar-source radiation, $H_{\rm *}$. 
As discussed below, because of the low interstellar medium (ISM) densities of our model galaxies, $H_{*}$ constitutes 
generally
the dominant heating term. The cosmic microwave background (CMB) provides
a temperature floor, $T_{\rm CMB}$, because it is thermodynamically
not possible to radiatively cool below.
Thus, the basic equation to be solved is:
\begin{equation}
{\Lambda}_{\rm d} (T_{\rm d}) - {\Lambda}_{\rm d} (T_{\rm CMB})
= H_{\rm d} + H_{\rm *} \mbox{\, .}
\label{eq:Td_eq}
\end{equation}
The procedure to estimate ${\Lambda}_{\rm d}$ and $H_{\rm d}$ is similar to that
in \citet{ji2014}, and we briefly describe it here for the convenience of
the reader, also noting that the
calculation of ${\Lambda}_{\rm d}$ closely follows the methodology in
\citet{schneider2006_methods}.

Dust grain emission is well approximated by thermal radiation
\citep{draine2001}, in which case the cooling rate can be written
\begin{equation} \label{eq:lambdad}
  \Lambda_{{\rm d}} (T_{\rm d}) = 4 \sigma_{{\rm SB}} T_{\rm d}^4 \kappa_{{\rm P,d}} \rho_{{\rm d}} \beta_{{\rm esc}} \mbox{\, ,}
\end{equation}
where $\sigma_{{\rm SB}}$ is the Stefan-Boltzmann constant,
$\kappa_{{\rm P,d}}$ the temperature-dependent Planck
mean opacity of dust grains per unit \emph{dust} mass,
$\rho_{{\rm d}}$ the dust mass density, and
$\beta_{{\rm esc}}$ the photon escape probability. For the
Planck dust opacities, we use the tabulated values derived by \citet{ji2014},
kindly provided by Alexander Ji (private communication). 
The escape fraction is estimated as $\beta_{{\rm esc}} = \min
(1,\tau^{-2})$, which is suitable for radiative diffusion out of an optically
thick gas cloud \citep{omukai2000}. The optical depth $\tau$ is given by:

\begin{equation} \label{eq:gasopacity} 
  \tau = (\kappa_{{\rm P,g}} {\rho}_{\rm g} + \kappa_{{\rm P,d}}\rho_{{\rm d}}) \lambda_{{\rm J}} \mbox{\, ,}
\end{equation}
where $\kappa_{{\rm P,g}}$ is the continuum Planck mean opacity of primordial gas
from \citet{mayer2005},
and $\lambda_{{\rm J}}$ the Jeans length.
The Jeans length is the typical size of a dense core
in a uniformly collapsing spherical gas cloud (e.g.
\citealt{larson1969}). For the densities encountered in this work, 
$\beta_{{\rm esc}} \approx 1$ provides a good approximation.

The gas-dust collisional heating rate \citep{hollenbach1979} is computed as:

\begin{equation}
H_{\rm d} = n n_{\rm d} {\sigma}_{\rm d} v_{\rm th} f (2 k_{\rm B} T_{\rm g} - 2 k_{\rm B} T_{\rm d}) \mbox{\, ,}
\label{eq:Hd}
\end{equation}

where $n$ is the number density of atomic hydrogen, $n_{\rm d}$
the number density of dust, $\sigma_{\rm d}$ the dust
geometrical cross section, $v_{{\rm th}}$ the thermal velocity of
atomic hydrogen, $T_{\rm g}$ the gas temperature, and
$f$ is a correction factor for species other than
atomic hydrogen.
Note that $n_{\rm d} {\sigma}_{\rm d} = {\rho}_{\rm d} S$, where $S$,
the total dust geometrical cross section per unit dust mass, is defined
as:

\begin{equation}
S = {\sum}_{i} f_i S_i \ {\rm with } \ S_i = \int_{0}^{\infty} \pi a^2 \frac{dn^i}{da}da \mbox{\, ,}
\end{equation}
with $f_i$ denoting the mass fraction, and the sum extending over the respective species considered in a given model.
Since higher energy particles collide more frequently \citep{draine2011},
the kinetic energy per colliding gas
particle is $2k_{\rm B}T$ instead of $1.5k_{\rm B}T$.
A Maxwellian velocity distribution is adopted for the gas so the
average velocity of atomic hydrogen is

\begin{equation}
  v_{\rm th} = \left(\frac{8k_{\rm B}T_{\rm g}}{\pi m_{\rm H}} \right)^{1/2}
\mbox{\, .}
\end{equation}

Finally, for obtaining the dust heating rate driven by stellar radiation,
we estimate the specific flux emerging from the central source, $f_{* , \nu}$,
that is absorbed by dust grains at a given radius $r$.
For simplicity, we assume isotropic emission and negligible extinction
between the stellar source and dust grains (optically thin medium). Hence,
\begin{equation}
\label{eq:fstarnu}
f_{* , \nu}  = \frac{L_{* , \nu}}{4 \pi r^2}, 
\end{equation}
where $L_{* , \nu}$ is the specific luminosity of the stellar cluster,
modelled as a point source.
The stellar luminosity is derived 
by rescaling Yggdrasil SEDs to the mass of the central cluster.
Thus,
\begin{equation}
H_{*} = {\int}_0^{\infty} {\kappa}_{\nu} {\rho}_{\rm d} f_{* , \nu} d{\nu}
\mbox{\, ,}
\label{eq:Hs}
\end{equation}
with ${\kappa}_{\nu}$ representing the frequency-dependent
dust opacity per unit mass calculated by \citet{ji2014}.  

Inside atomic cooling haloes, $T_{\rm g}$ is expected to range between 
$\sim 10^2$ and $10^4$ K, depending on the physical condition of the gas phase
\citep[e.g.][]{safranek2010, safranek2014}.  
However, because of the low density
of the gas component encountered in this work, our results indicate that
$T_{\rm d}$ is not very sensitive to variations in $T_{\rm g}$.
We verified this for all systems considered here, specifically for
$M_{\rm vir} \ge 10^7 M_{\odot}$ and $z \ \epsilon \ [7, 20]$,
by assuming an isothermal gas component with three 
different values of $T_{\rm g}$: $10^2 , 10^3, 10^4$ K. We obtained, in all cases, similar
values of $T_{\rm d}$ because 
the stellar heating rate dominates 
over that driven by gas-dust collisions: $H_{\rm d} / H_* < 10^{-6}, 10^{-5},
10^{-3}$ for $T_{\rm g} = 10^2 , 10^3, 10^4$ K, respectively.  
One can similarly gauge the unimportance of the collisional heating term
by estimating the typical timescale for
gas-grain collisions: 
$t_{\rm coll} > 3 \times 10^5 {\rm s}, 10^5 {\rm s}, 3 \times 10^4 {\rm s}$
for $T_{\rm g} = 10^2 , 10^3, 10^4$ K, respectively. These are long
compared to the timescale for radiative heating. Thus, since the detailed
gas properties have a non significant effect on our estimates, for the sake of simplicity,
we assume $T_{\rm g} = 10^3$\,K throughout. We also adopt $f = 1$, 
corresponding to atomic hydrogen. 

\begin{figure}
\begin{center}
\resizebox{7.5cm}{!}{\includegraphics{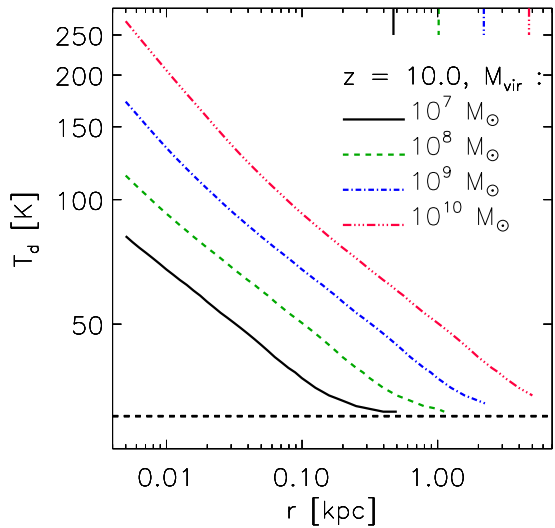}}
\end{center}
\caption[Td vs r at $z=10$]
{
Dust temperature profiles for the same systems shown in Fig. \ref{fig:ng_vs_r_z10},
generated by assuming the UM-ND-20 dust model and standard size distribution 
for dust grains (see text for details).
The vertical lines on the top again depict the virial radii of these systems.
The horizontal dashed line denotes the CMB temperature at $z=10$.
Note that temperatures are higher in the more massive systems, since
the luminosity of the central star cluster is scaled up accordingly. 
}
\label{fig:Td_vs_r_z10}
\end{figure}

As an example, in Fig. \ref{fig:Td_vs_r_z10}, we present the dust
temperature profile corresponding to the same systems shown in Fig. \ref{fig:ng_vs_r_z10}, where the
UM-ND-20 model and standard grain size distribution has been assumed.
Note that $T_{\rm d}$ increases significantly as $r \rightarrow 0$ because of
the divergence of $f_{*, \nu}$ (Equ. \ref{eq:fstarnu}). Therefore, in order to avoid
numerical artefacts, we impose a lower cut-off for the radius at $r_{\rm cut} = 5\ {\rm pc}$,
as this is the typical size of a stellar cluster \citep[e.g.][]{safranek2014}.
Furthermore, when $T_{\rm d}$ becomes sufficiently high, dust sublimates. The sublimation
temperature depends on dust composition. Here, we again follow \citet{ji2014} and assume
a sublimation temperature of 1500\,K, typical of non-carbon grains \citep{schneider2006_methods}.
In our model, central dust temperatures, at $r = r_{\rm cut}$, increase with halo mass,
and mainly depend on the size distribution of dust grains. For the standard distribution,
the maximum $T_{\rm d}$ ranges from $\approx 70$\,K ($M_{\rm vir} \approx 10^7 M_{\sun}$) 
to 1500\,K ($M_{\rm vir} \gtrsim 10^{12} M_{\sun}$), whereas for the shock size distribution,
this range extends from $\approx 100$\,K ($M_{\rm vir} \approx 10^7 M_{\sun}$)
to 1500\,K ($M_{\rm vir} \gtrsim 10^{11} M_{\sun}$).
Because of sublimation, dust is suppressed at $r \sim r_{\rm cut}$ in more massive
galaxies ($M_{\rm vir} \gtrsim 10^{11-12} M_{\sun}$).  
As a baseline reference, for a typical atomic cooling halo 
($M_{\rm vir} \approx 10^8 M_{\sun}$), the maximum $T_{\rm d}$ lies 
between 100\--120\,K (standard size distribution) and 150\--180\,K (shock size distribution),
depending on the dust composition.
It is worth noting that, for carbon dust models, the sublimation temperature would increase to 2000\,K,
thus allowing higher central temperatures in more massive systems.

\subsubsection{Dust emission}
\label{sec:dust_emission}

We compute the dust emissivity per unit mass ($j_{\nu}$), at a given radius, by 
applying Kirchhoff's law for the estimated $T_{\rm d}$ profile:
\begin{equation}
j_{\nu} ( T_{\rm d} ) = 4 \pi {\kappa}_{\nu} B_{\nu} (T_{\rm d}),
\end{equation}
where $B_{\nu}$ is the Planck function.
The total specific {\em dust} luminosity $L_{\nu ,{\rm em}}$ emitted 
by the system is obtained by integrating $j_{\nu}$ out to $R_{\rm vir}$:
\begin{equation}
L_{{\nu}, {\rm em}} = 4 \pi \int_{r_{\rm cut}}^{R_{\rm vir}} {\rho}_{\rm d} (r) j_{\nu} (r) r^2 dr \mbox{\, .}
\label{eq:L_nu}
\end{equation}
The observed {\em dust} specific flux $f_{\nu , {\rm obs}}$ originating from the model galaxy is then:
\begin{equation}
f_{\nu , {\rm obs}} = (1 + z)  \frac{L_{\nu ,{\rm em}}}{4 \pi {d_{L}}^2}
\mbox{\, ,}
\end{equation}
where $d_{L}$ is the luminosity distance to a source at redshift $z$.

\begin{figure}
\begin{center}
\resizebox{7.5cm}{!}{\includegraphics{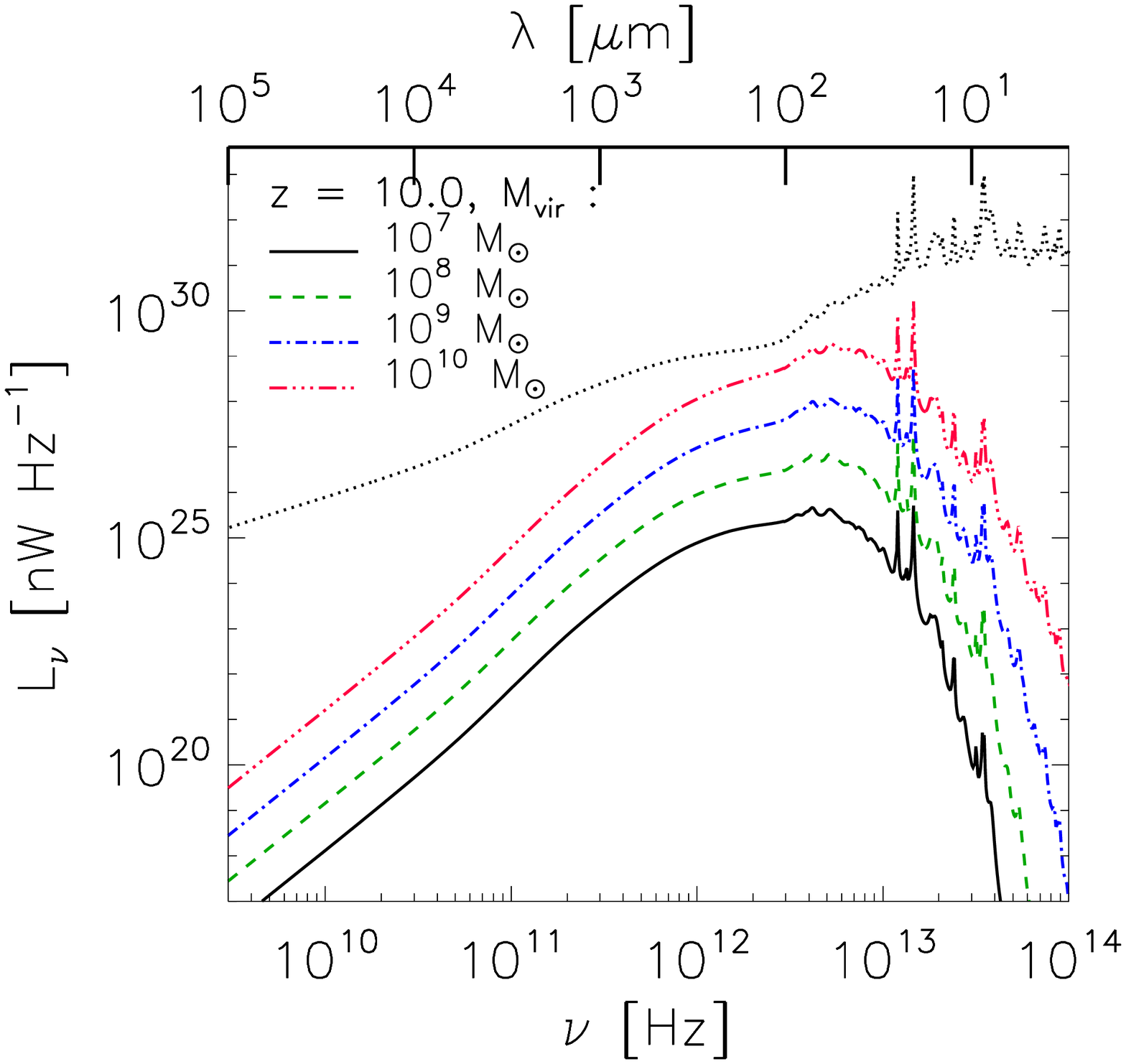}}\\
\resizebox{7.5cm}{!}{\includegraphics{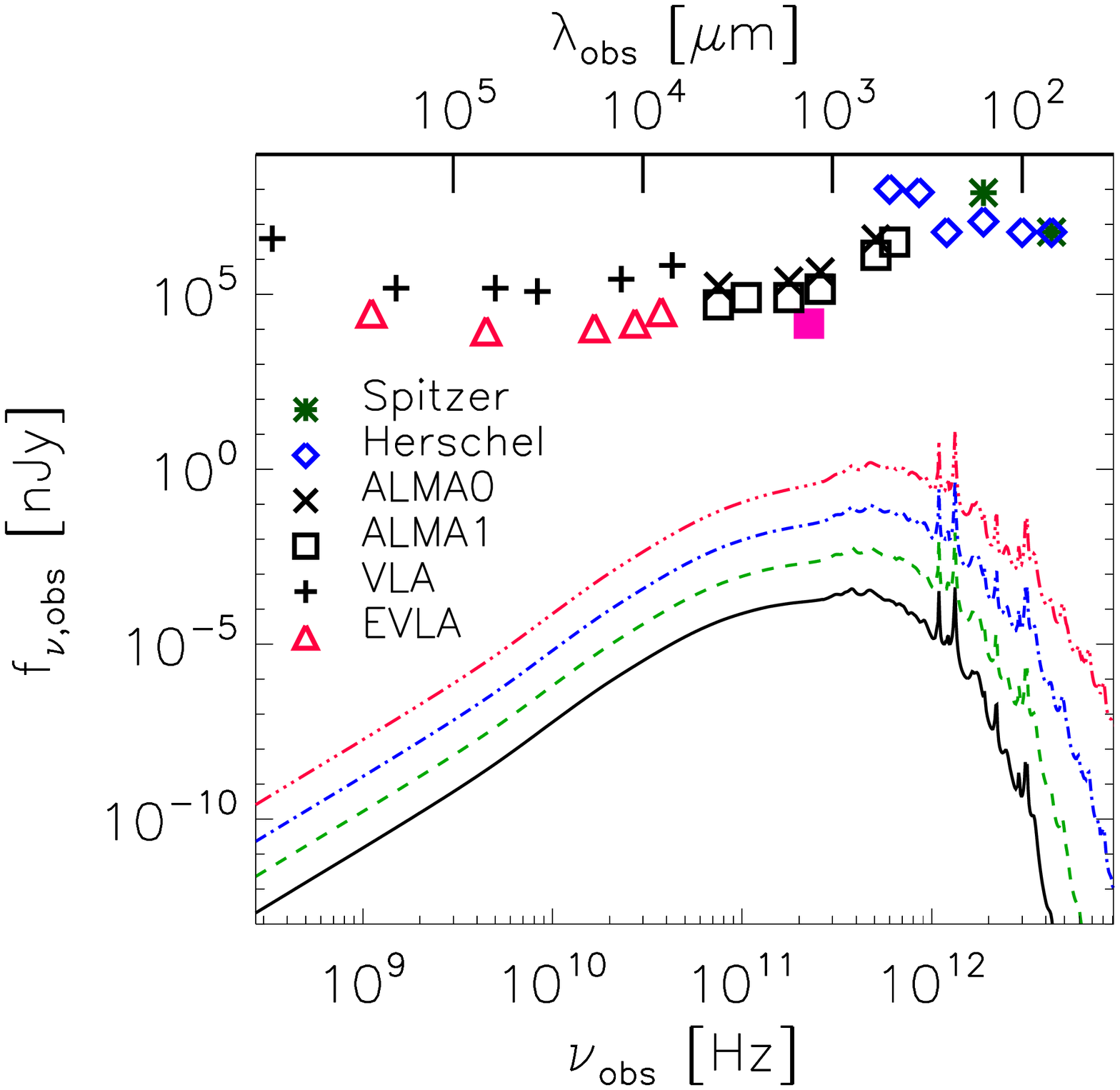}}
\end{center}
\caption[galaxy spectra at $z=10$]
{Dust re-emission spectra for individual sources.
{\it Upper panel:} Rest-frame spectra
for the same systems shown in Fig.~\ref{fig:ng_vs_r_z10}, as indicated
in the figure. The UM-ND-20 model and standard grain-size distribution
has been assumed.
The dotted black curve depicts the frequency-dependent
dust opacity ${\kappa}_{\nu}$, which has been re-scaled arbitrarily for the sake of comparison. 
{\it Lower panel:} Observed specific flux.
Sensitivities of different instruments are shown with symbols as indicated
in the figure (see text for details).
The pink solid square represent one of the highest sensitivities achieved
for ALMA maps in recent works \citep[$\sim 10 \mu {\rm Jy / beam}$, e.g.][]{carniani2015}.
Note that we are only dealing with {\em dust} emission in this work 
so that the contribution from stellar radiation to the total luminosity 
is not included in the calculations.
}
\label{fig:spectra_z10}
\end{figure}

In Figure~\ref{fig:spectra_z10}, we show the emitted spectra (upper panel), together with the observed
specific flux (lower panel), corresponding to the systems in Fig.~\ref{fig:ng_vs_r_z10}
for the UM-ND-20 model and a standard size distribution for dust grains.
Note that these spectra only represent {\em dust} emission; thus, they do 
not include the contribution from stellar sources.
The dotted black curve, in the upper panel, depicts the frequency-dependent
dust opacity, ${\kappa}_{\nu}$, which has been re-scaled arbitrarily for the sake of comparison.
The maximum emission is obtained at $\lambda \sim 50 \mu {\rm m}$, corresponding to
an observed wavelength of ${\lambda}_{\rm obs} \sim 500 \mu {\rm m}$. At longer
wavelengths, the spectral slope is determined by the Rayleigh-Jeans tail of the 
distribution ($B_{\nu} \sim {\nu}^2$), modulated by the frequency-dependent dust opacity 
(${\kappa}_{\nu} \sim {\nu}^{1.6}$, for the UM-ND-20 dust model). Thus, 
$f_{{\nu},{\rm obs}} \sim {\nu}_{\rm obs}^{3.6}$ over this wavelength regime.
At shorter wavelengths, the spectral shape traces the different absorption features
associated with the given dust model (dotted black curve).

In the lower panel of Fig.~\ref{fig:spectra_z10}, we additionally show the sensitivities of
different observatories: Spitzer (green asterisks), Herschel (blue diamonds), 
ALMA (black X and squares for cycles 0 and 1 observations, respectively),
the Very Large Array (VLA, black crosses), and the Expanded Very Large Array (EVLA, red triangles).
\footnote{Sensitivities were taken from the {\em James Webb Space Telescope} website
at http://www.stsci.edu/jswt/science/sensitivity, and correspond to the faintest flux
for a point source that can be detected at SNR$=10$ in a $10^4$ s integration.
We refer the reader to this web-page for further details.}
The pink solid square represent one of the highest sensitivities achieved
for recent ALMA maps \citep[$\sim 10 \mu {\rm Jy / beam}$, e.g.][]{carniani2015}. 
We see that the peak of the predicted dust emission is
about four orders of magnitude below instrumental capabilities. 
Accordingly, point source sensitivities of current and near-future instruments
are not sufficient to allow detection of dust emission from individual dwarf-size galaxies in the early Universe.

Confusion noise may affect the detection of luminous systems independently of instrument noise. 
We perfomed a rough estimate of confusion for our model galaxies following the procedure
described in Section~{4.1} in \citet{kashlinsky2015}. 
We obtained $\lesssim 10$ and $\lesssim 0.003$ beams/source for sources at $z < 20$ 
considering beam sizes
of 0.7" and 37", respectively, consistent with the spatial resolution of ALMA compact configuration and  
Herschel at $\lambda \sim 500 \ {\rm \mu m}$.
Assuming that confusion intervenes when there are less than 50 beams/source \citep{condon1974}, 
all our sources are well within the confusion of ALMA compact configuration and Herschel
instruments.  However, for ALMA more extended 12-m array configurations, the finest angular resolutions 
for cycles 3 and 4 observations reach 0.04"-0.032". We obtained that, in the latter case,
model sources at $z > 13$ could overcome the confusion but, as we have shown, 
they would be too faint to be detected by ALMA.

As emission tends to significantly increase with mass, rare massive
systems might be detectable, but are statistically difficult to find. 
In fact, by extrapolating our model to higher masses, we find that a halo mass
of $M_{\rm vir} \ge 10^{14} - 10^{15} M_{\sun}$ would be required to reach instrumental
sensitivities. Such an object would have a {\em dust} luminosity of at least
$L_{\rm em} \sim 10^{12} - 10^{13} L_{\sun}$, consistent with
the prototypical local ultra-luminous starburst galaxy
Arp 220 ($\sim 10^{12} L_{\sun}$). Note, however, that at these high masses
additional physical processes might have to be considered to
correctly model dust emission, such as 
active galactic nuclei heating and the prolific dust generation after
merger-induced starbursts.

Finally, we have assumed a conservative dust-to-metal ratio
of $D = 0.005$. For larger $D$, the predicted
dust emission would increase almost proportionally (Equ.~\ref{eq:L_nu}), and the
probability of observing these sources would be higher. 
As mentioned above, we show results for the UM-ND-20 dust model combined with a standard size
distribution for dust grains. Other dust chemical compositions
lead to similar findings. On the other hand, when using the shock size distribution, 
the predicted luminosities and fluxes increase by a factor $\lesssim 10$, where the exact
boost depends on the dust chemistry (see also Sec.~\ref{sec:CIB}).
Another assumption made here is a negligible contribution of carbon to the dust 
chemical composition. \citet{schneider2006_methods} compared opacities derived from a silicon-dominated
dust model and one with similar contributions from silicon and carbon (their figures 1 and 2).
We expect that our dust temperature profile does not exhibit significant variations
if a moderate amount of carbon were included: the steep temperature
dependence of the dust cooling rate (Equ.~\ref{eq:lambdad}) implies that 
opacity differences between models could be compensated by slight variations in $T_{\rm d}$.
However, at a given $T_{\rm d}$, dust emissivities would be a factor of a few higher if
we had adopted the \citet{schneider2006_methods} model that includes comparable amounts of silicon
and carbon. Thus, a carbon-based model combined with a shock size distribution could increase the dust
emission by a factor of $\sim 10$, which still remains below instrumental capabilities, but
a more detailed analysis is beyond the scope of the present paper.

It is worth mentioning that observations by \citet{leiton2015} suggest that
the main contributors to the EBL in all Herschel bands seem to be distant siblings 
of the Milky Way ($z \sim 1.0$ for $\lambda <300 \mu {\rm m}$) with a stellar mass of 
$M_{*} \sim 9 \times 10^{10} M_{\sun}$.
Also, \citet{viero2013} suggest that the EBL is dominated by systems with
$M_* \sim 10^{9.5 - 11} M_{\odot}$, and that the sources associated with wavelengths
below and above $200 \mu {\rm m}$ are located at $z<1$ and $1<z<2$, respectively.
The low fluxes predicted by our models for individual primeval dust sources are consistent with these empirical constraints.

In the following section, we analyse the possibility of detecting signatures of 
integrated dust emission by studying the contribution 
of model galaxies to the EBL.

\section{The cosmic FIR background}
\label{sec:CIB}

In order to estimate the contribution of the first galaxies to 
the FIR/sub-mm cosmic background,
we combine the Sheth-Tormen mass function
\citep{sheth2001} with our idealized models for individual sources
in Sec.~\ref{sec:dust_model}.
As explained above, our models allow us to estimate the
dust emission from primeval systems, once their properties are specified,
such as their stellar, gas and dust distributions.

For sources located in a given redshift range $[z_{\rm min}, z_{\rm max}]$, 
the specific intensity contributed to the cosmic background at observed frequency ${\nu}_{\rm obs}$
is given by:

\begin{equation}
\label{eq:Inu}
I_{\nu} ({\nu}_{\rm obs}) = \frac{c}{4 \pi} \int_{z_{\rm min}}^{z_{\rm max}} 
{\epsilon}_{\nu}(\nu , z) \ \left| \frac{{\rm d}t}{{\rm d}z} \right| \ {\rm d}z
\end{equation}

where ${\epsilon}_{\nu}(\nu , z)$ is the specific luminosity per comoving volume element
at redshift $z$ and ${{\nu} = \nu}_{\rm obs} (1 + z)$ is the frequency at the rest frame
of the source. Standard cosmology gives the expression:

\begin{equation}
\left| \frac{{\rm d}t}{{\rm d}z} \right|^{-1} = H_0 \ (1 + z) \
[ (1 + z)^2 (1 + {\Omega}_{\rm m} z) -
z (2 + z) {\Omega}_{\Lambda} ]^{1/2} 
\end{equation}

We estimate ${\epsilon}_{\nu}(\nu , z)$, associated to dust emission, by combining
the Sheth-Tormen mass function $n_{\rm ST} (M_{\rm vir}, z)$
\footnote{The Sheth-Tormen formalism assumes ellipsoidal instead of
spherical collapse and provides a better fit to simulations
than the Press-Schechter (PS) formalism. The PS mass function
underpredicts the abundance of haloes at high redshifts.}
 with the specific {\em dust} luminosities
$L_{\nu} (M_{\rm vir}, z)$ obtained by our idealized models:

\begin{equation}
\label{eq:epsilon_nu_z}
{\epsilon}_{\nu}(\nu , z) =  \int_{M_{\rm min}}^{M_{\rm max}}
L_{\nu, {\rm em}} (M_{\rm vir}, z) \ n_{\rm ST} (M_{\rm vir}, z) \ {\rm d}M_{\rm vir}
\end{equation}
We here consider sources with typical masses and redshifts associated with
first galactic systems: $M_{\rm vir} \ge 10^7 M_{\odot}$ and  $z \ \epsilon \ [7, 20]$.

\begin{figure*}
\begin{center}
\resizebox{7.5cm}{!}{\includegraphics{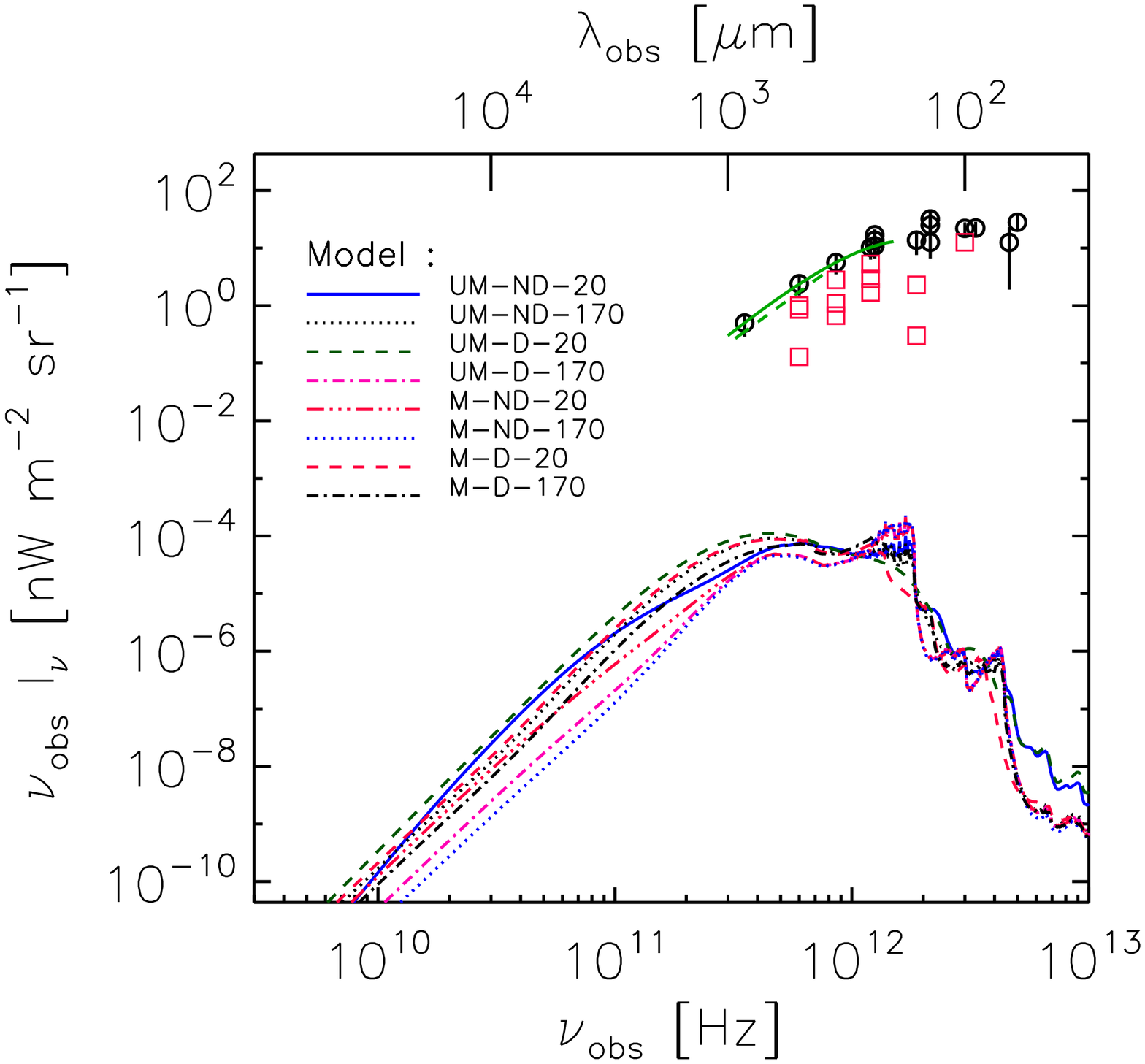}}
\resizebox{7.5cm}{!}{\includegraphics{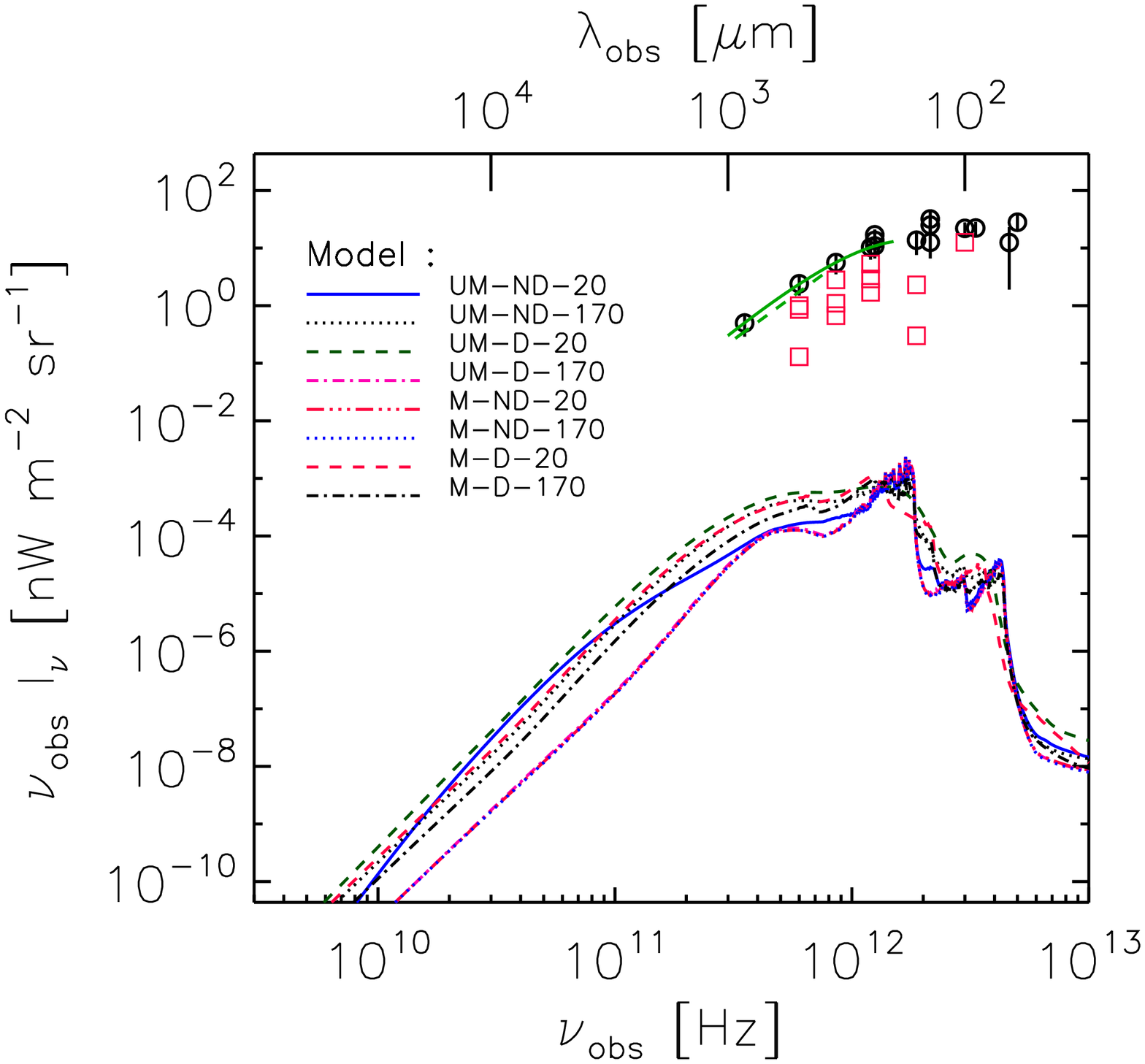}}
\end{center}
\caption[Background]
{FIR/sub-mm-EBL from the first galaxies. We show the
specific intensity, $I_{\nu}$, of the cosmic background at observed frequency ${\nu}_{\rm obs}$,
considering sources with $M_{\rm vir} \ge 10^7 M_{\odot}$ located at $z \ \epsilon \ [7, 20]$.
{\it Left panel:} Results obtained by using the standard size distribution for dust grains.
{\it Right panel:} Results for shock size distribution.
Different dust models are considered as indicated in the figure.
The light-green curves and circles with error bars depict observed measurements of the cosmic 
background derived from {\em Akari}, COBE/DIRBE and COBE/FIRAS instruments. Red squares
represent an estimation of the CIB excess after removing the contribution 
of the IGL obtained by stacking analysis (see text for details). 
}
\label{fig:background}
\end{figure*}

In Figure~\ref{fig:background}, we present the far-IR/sub-mm cosmic background predicted
by our model assuming different dust compositions
and grain size distributions. Dust emission from the first galaxies 
contributes most significantly to the  EBL at
${\lambda}_{\rm obs} \approx 100{\--}1000 \mu {\rm m}$, exhibiting
a peak at $\sim 500 \mu {\rm m}$.
The spectral slope at longer wavelengths is given by the Rayleigh-Jeans law ($\sim {\nu}^2$)
combined with the frequency-dependent dust opacity (${\kappa}_{\nu} \sim {\nu}^{n}$, with
$n$ ranging from 1.0 to 1.8, depending on the dust model). Hence, in the long-wavelength regime,
${\nu_{\rm obs}} I_{\nu} \propto {\nu}_{\rm obs}^{m}$, with $m$ between 4.0 and 4.8.
At shorter wavelengths, the spectral shape traces the different features associated
with the different dust opacities. 
These spectral features can also be discerned in Fig.~\ref{fig:spectra_z10}, but
in Fig.~\ref{fig:background} they are smoothed and amplified because of the 
cumulative effect of sources contributing from different $z$ (Equ.~\ref{eq:Inu}).
For example, the spectral break at 
${\lambda}_{\rm obs} \approx 200  \mu {\rm m}$ in Fig.~\ref{fig:background}
corresponds to the cumulative,
redshifted feature in ${\kappa}_{\nu}$ at $\nu \sim 10^{13}$ Hz (see Fig.~\ref{fig:spectra_z10},
upper panel).

It is evident that the dust chemical composition does not significantly affect
the main trends of the background radiation. 
On the other hand, changing the size distribution of dust grains seems
to be more significant. 
In the case of the shock size distribution, the predicted background is around one
order of magnitude higher than that corresponding to the standard size distribution.
The absolute maximum intensities are $\sim 10^{-4} {\rm nW \ m^{-2} \ sr^{-1}}$ and
$\sim 10^{-3} {\rm nW \ m^{-2} \ sr^{-1}}$ for the standard and shock size distributions,
respectively.

The light-green lines and black circles in Fig.~\ref{fig:background} show 
absolute measurements of, and limits on, the extragalactic background light taken from
tables $3-5$ in \citet{dwek2012}, based on data obtained by
different methods with an array of satellites, balloon-experiments,
and ground-based observatories. 
At ${\lambda}_{\rm obs} = 65$ and $90 \ {\mu}{\rm m}$,
data from the {\rm Akari} infrared imaging satellite (ASTRO-F) 
has been used \citep{matsuura2011}.
Measurements from the Diffuse Infrared Background Experiment (DIRBE) on board 
the COsmic Background Explorer (COBE) satellite are shown at 
${\lambda}_{\rm obs} = 60 \ {\mu}{\rm m}$ \citep{finkbeiner2000},
$100 \ {\mu}{\rm m}$ \citep{hauser1998} and $140 \ {\mu}{\rm m}$ \citep{hauser1998, schlegel1998}.
Results obtained by \citet{fixsen1998} with COBE's Far Infrared Absolute Photometer (FIRAS) Pass 4 
are plotted at ${\lambda}_{\rm obs} = 140, 160, 240, 250, 350, 500$ 
and $850 \ {\mu}{\rm m}$.
The solid light-green line corresponds to the analytical fit to the spectrum given by
\citet{fixsen1998}, while the dashed light-green line denotes the tentative background derived by
\citet{puget1996} from COBE/FIRAS Pass 3.
In addition, we perform a rough estimation of the average source-subtracted
EBL (red squares in Fig.~\ref{fig:background}) by removing the integrated galactic light (IGL) 
associated with foreground sources from the absolute EBL measurements.
For this calculation, we used absolute measurements reported by \citet{hauser1998} 
at ${\lambda}_{\rm obs} = 100 {\mu}{\rm m}$ and \citet{fixsen1998} at longer wavelengths.
To estimate the IGL, we employed results derived by different authors 
from stacking analysis of astronomical images.
Data at ${\lambda}_{\rm obs} = 100 - 160 \ {\mu}{\rm m}$
were obtained from the Photodetector Array Camera (PACS) on board Herschel 
\citep{dole2006, berta2010}. At ${\lambda}_{\rm obs} > 200 \ {\mu}{\rm m}$,
we used measurements from the Balloon-borne Large-Aperture Submillimeter Telescope 
(BLAST; \citealt{marsden2009, bethermin2010}) and the 
Spectral and Photometric Imaging Receiver (SPIRE) on board Herschel 
\citep{bethermin2012}. 
It is worth noting the large uncertainties ($\sim 1 - 10 {\rm \ nW \ m^{-2} \ sr^{-1}}$)
involved in the estimation of the source-subtracted EBL component, so caution should be taken
when drawing conclusions from these data.

As can be seen, our model predictions fall below the measured absolute background by $\sim 3 - 4$ 
orders of magnitude, depending on the dust model.
Moreover, our model spectrum is also below the average source-subtracted EBL by $\sim 2 - 3$
orders of magnitude.
Thus, dust emission from the first galaxies might not
contribute significantly to the observed absolute cosmic far\--IR/sub\--mm background.
If this is the case, the bulk of the observed radiation in the FIR-EBL
could be related to more evolved massive galaxies, located at lower
$z$.
It is worth mentioning that our model can roughly reproduce
the Rayleigh-Jeans tail of the observed distribution. This suggests that
the sources that contribute the bulk of the observed radiation should exhibit, on average, similar
dust opacity trends as those assumed here.

\begin{figure*}
\begin{center}
\resizebox{17.5cm}{!}{\includegraphics{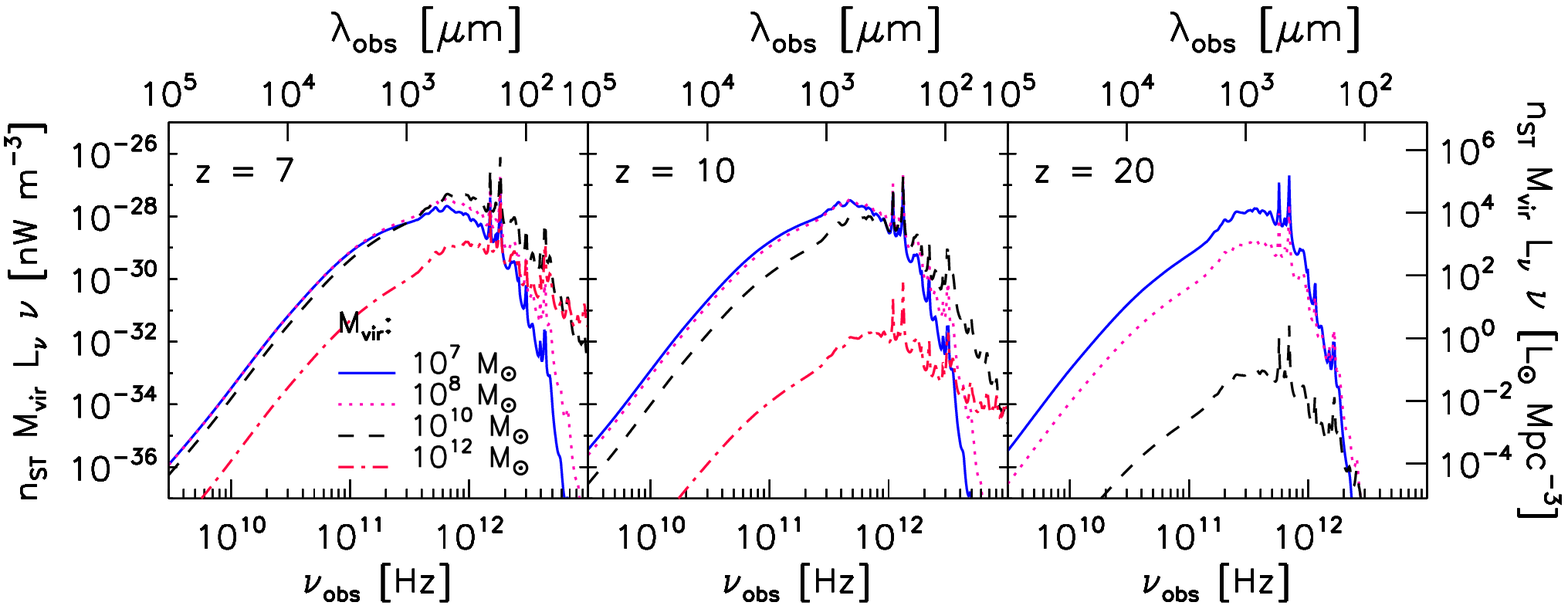}}\\
\resizebox{17.5cm}{!}{\includegraphics{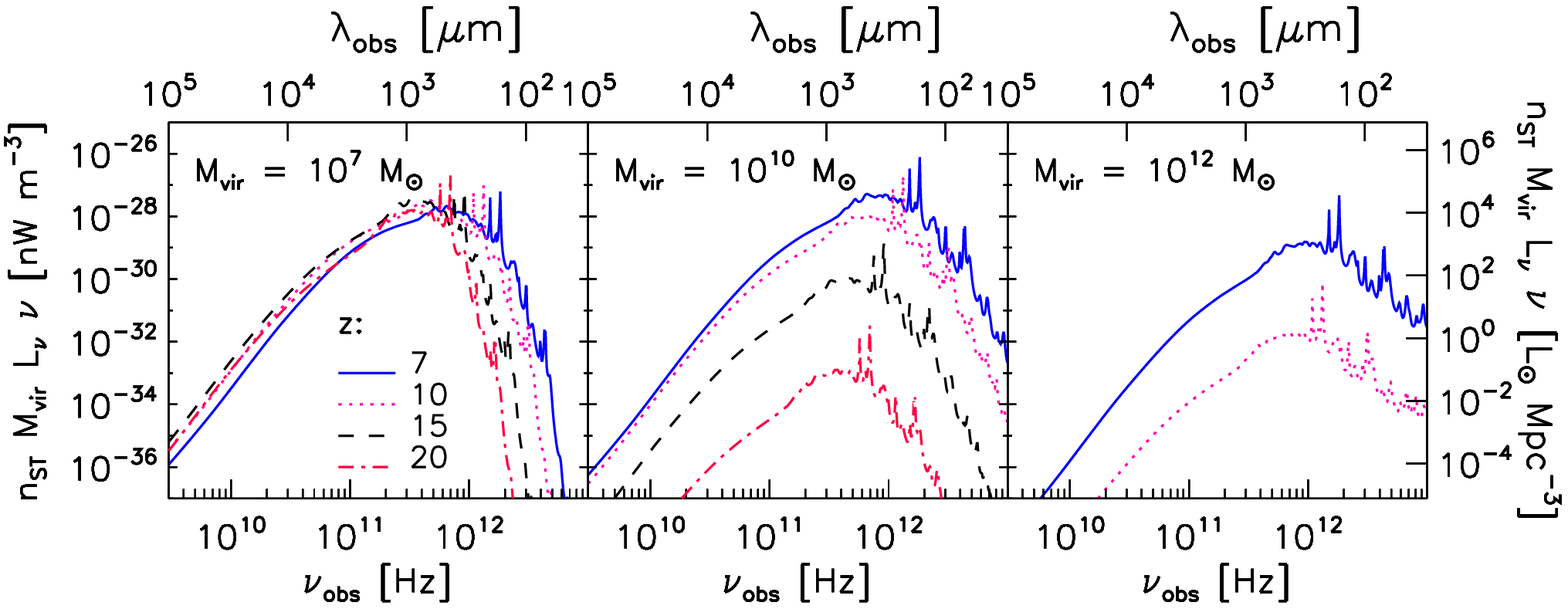}}
\end{center}
\caption[nLM]
{
Energy emitted per unit time and comoving volume by model galaxies as a function
of ${\nu}_{\rm obs}$.
Upper panels compare the power emitted by systems of different masses at a fixed $z$, 
while lower panels show the energy contribution from systems of a given mass at 
different $z$. Left and right vertical axis show results in ${\rm nW \ m^{-3}}$ 
and ${\rm L_{\sun} \ Mpc^{-3}}$, respectively. 
For the sake of clarity, results are shown for the UM-ND-20 dust model combined
with a standard size distribution for dust grains. We have verified that other dust models produce
similar trends. 
}
\label{fig:nLM}
\end{figure*}

To ascertain the nature of the sources contributing most significantly to the
FIR-EBL, in Fig. \ref{fig:nLM}, we analyse the luminosity density generated by model galaxies
of different masses as a function of $z$.
Results are shown for the UM-ND-20 dust model combined
with a standard size distribution for dust grains. Other dust models produce
similar trends.
In the upper panels, we compare the luminosity density emitted by systems of different masses at a given $z$.
It is evident that radiation from smaller galaxies ($M_{\rm vir} \lesssim 10^{10} M_{\sun}$) dominates the
EBL at $\lambda_{\rm obs} \gtrsim 500 \mu {\rm m}$ over the entire redshift range
($z = 7 - 20$). On the other hand, at $z \lesssim 10$, 
larger galaxies ($M_{\rm vir} \gtrsim 10^{10} M_{\sun}$) become the dominant contributors
to the EBL at $\lambda_{\rm obs} \lesssim 500 \mu {\rm m}$.
Similarly, in the lower panels of Fig. \ref{fig:nLM}, we show the energy contribution from systems of a fixed mass at
different $z$. We can see that massive galaxies ($M_{\rm vir} \gtrsim 10^{10} M_{\sun}$) significantly contribute
to the spectra only at $z \lesssim 10$, because of the increase in the number density
of massive haloes at lower $z$. At the longest wavelengths ($\lambda \gtrsim 500 \mu {\rm m}$),
the main contributors to the EBL at $z \gtrsim 10$ are
small galaxies ($M_{\rm vir} \sim 10^{7 \-- 8} M_{\sun}$).
To sum up, according to our model, the bulk of the FIR/sub\--mm EBL originates in dwarf galaxies ($M_{\rm vir} \lesssim 10^{10} M_{\sun}$) at $z= 7\--20$, while radiation at shorter wavelengths can be associated with more massive systems 
at $z \sim 7$.

\subsection{Sensitivity to variations of model parameters}

In the following, we explore the sensitivity of our results to four
of our most critical model parameters:
the density profile, the dust-to-metal ratio $D$, the ISM metallicity $Z_{\rm g}$ and
the star formation efficiency $\eta$.

With respect to the gas density, we have used an isothermal 
1D profile (Equ.~\ref{eq:rhog_vs_r}). 
In addition, we have tested the effects of using a Navarro-Frenk-White (NFW)
density profile \citep[][]{navarro1997} outside the core radius $R_{\rm c}$:

\begin{equation}
\label{eq:rhog_vs_r_NFW}
{\rho}_{\rm g} (r \ge R_{\rm c}) = \frac{{\rho}_{0}}{\frac{r}{r_0} \  \Big(1 +  \frac{r}{r_0}\Big)^2}
\end{equation}

We have also
implemented a profile of the \citet{burkert1995} form:

\begin{equation}
\label{eq:rhog_vs_r_b}
{\rho}_{\rm g} (r) = \frac{{\rho}_{0} r_{0}^3}{(r + r_0 ) (r^2 + r_{0}^2)}
\end{equation}

In both cases, the parameters ${\rho}_{0}$ and  $r_0$ 
were chosen to obtain similar central densities and total gas masses
as those derived from Equ.~\ref{eq:rhog_vs_r}. 
In comparison with the isothermal density profile, the implementation of the NFW profile 
at $r \ge R_{\rm c}$ generates a mild decrease of mass at large radii and, consequently, 
an increase of it at intermediate radii.
On the other hand, the use of the \citet{burkert1995} profile, instead of the isothermal one,
results in a mild decrease of the core mass, a mild increase of
the gas and dust masses at intermediate radii and a moderate decrease of mass density towards $R_{\rm vir}$. 
We found that neither of these changes do
significantly alter our previous results, because
$H_{\rm d} / H_{*} \ll 1$ and ${\beta }_{\rm esc} \approx 1$. Thus,
the $T_{\rm d}$ profile is not very sensitive to variations
in ${\rho}_{\rm d}$ (Equ.~\ref{eq:Td_eq}).

\begin{figure*}
\begin{center}
\resizebox{7.5cm}{!}{\includegraphics{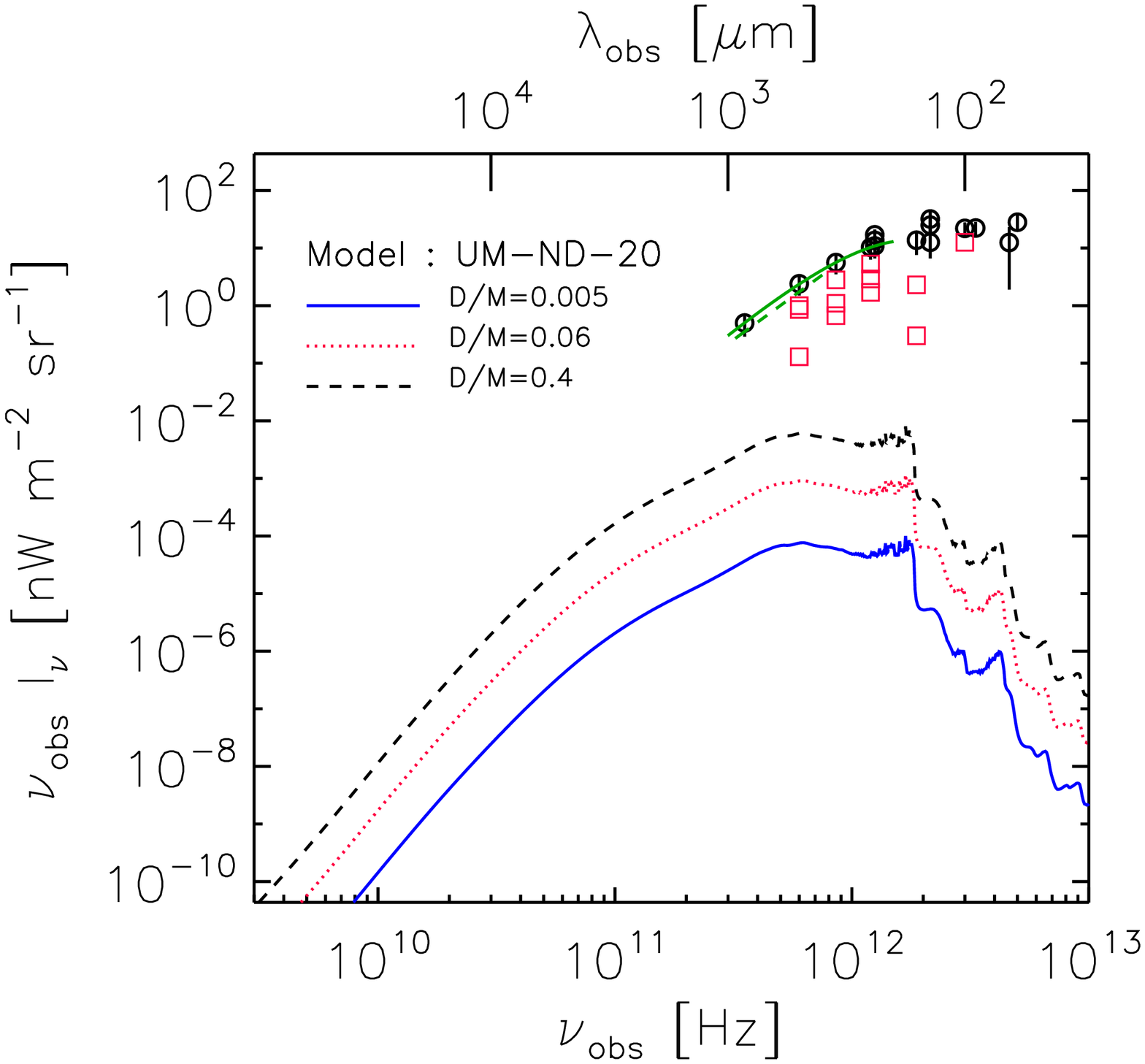}}
\resizebox{7.5cm}{!}{\includegraphics{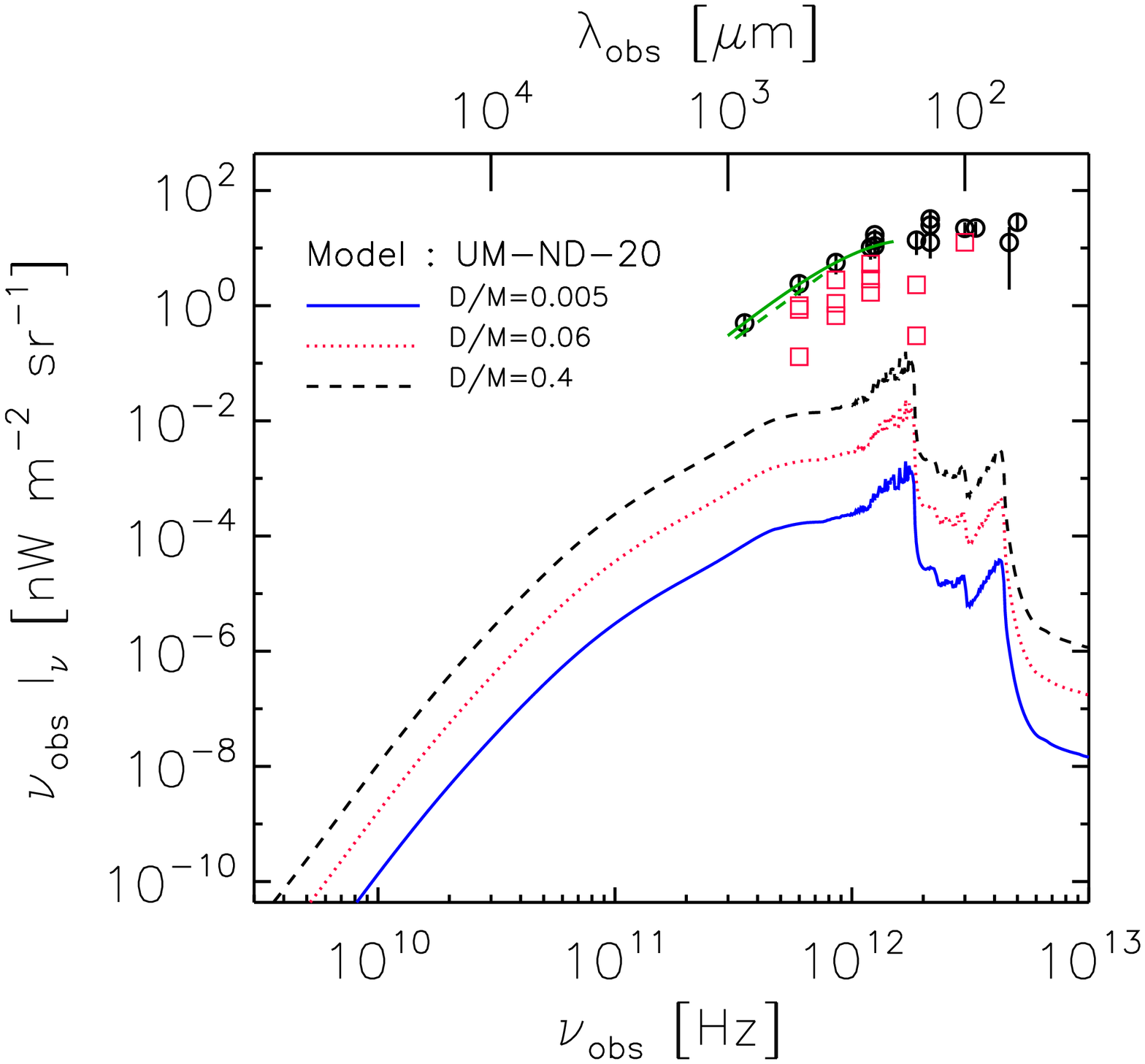}}\\
\vspace{-0.5cm}
\resizebox{7.5cm}{!}{\includegraphics{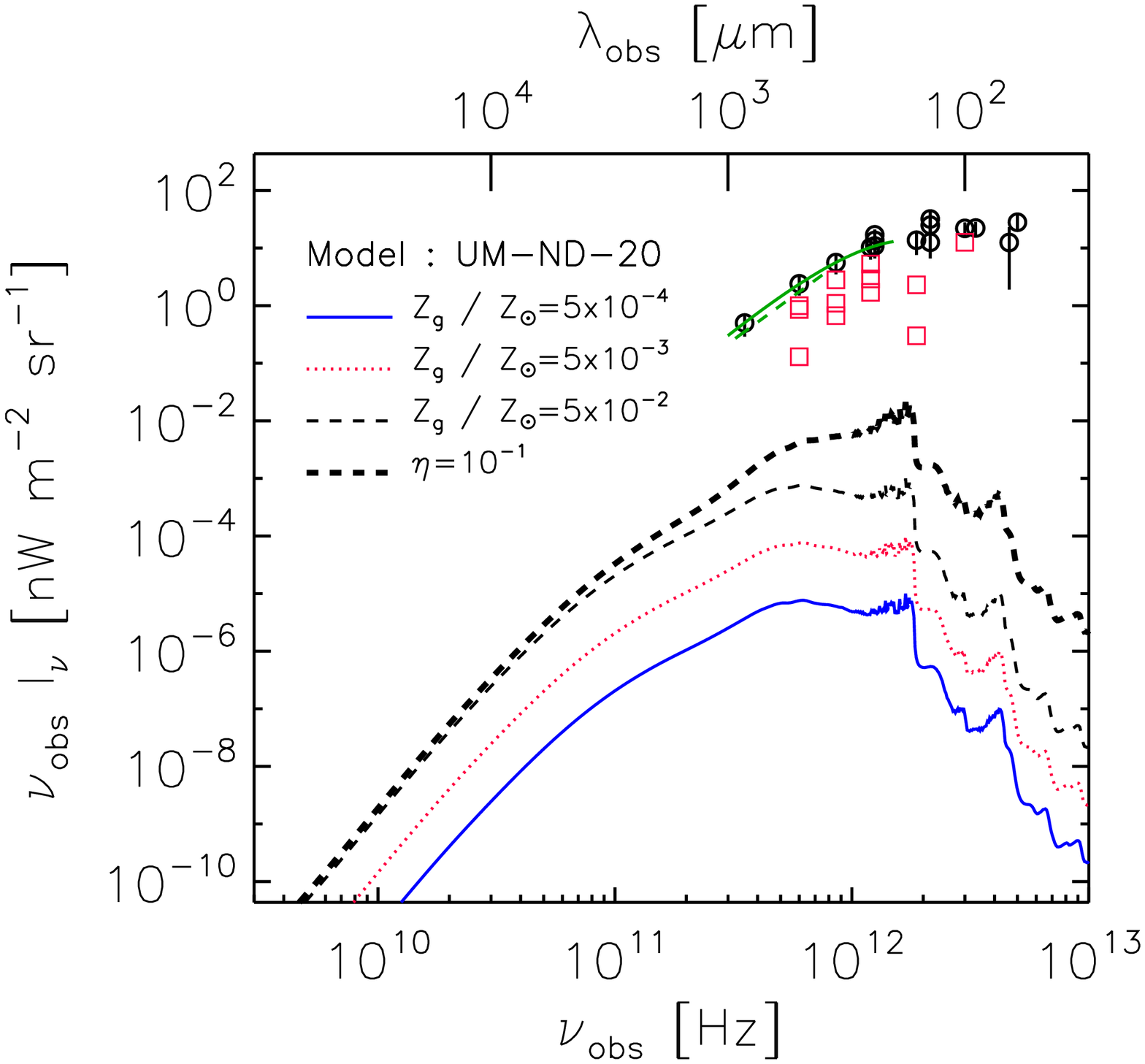}}
\resizebox{7.5cm}{!}{\includegraphics{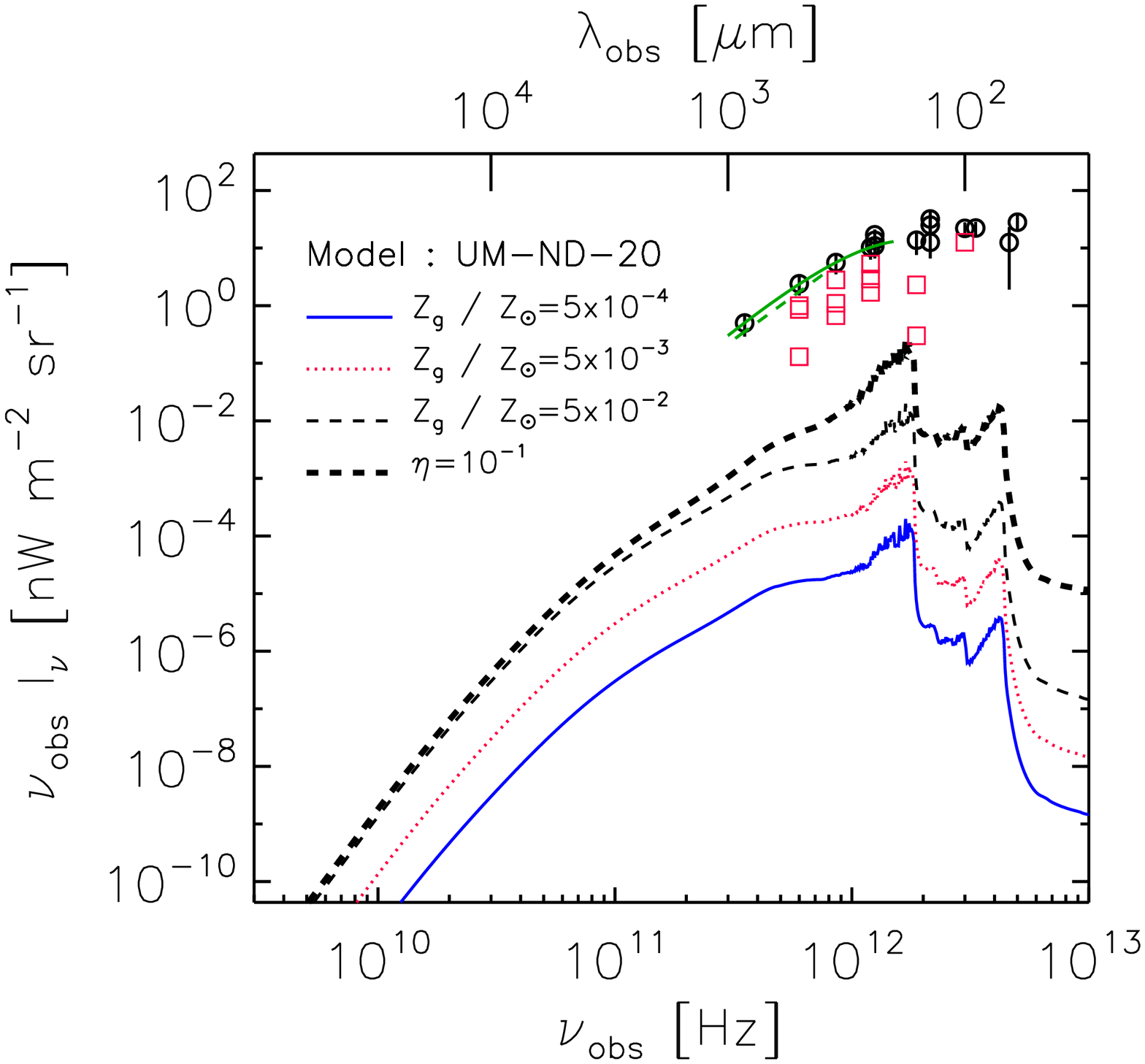}}
\end{center}
\caption[Background]
{
Similar to Fig. \ref{fig:background} but showing the effect
of changing the dust-to-metal ratio (upper panels) and metallicity (lower panels)
 for a given dust model (UM-ND-20, in this case).
In the lower panels, the thicker black dashed line corresponds to 
$Z_{\rm g}= 10^{-2} \ Z_{\sun}$ and $\eta = 10^{-1}$.
Left panels show results obtained by using the standard size distribution for dust grains,
while right panels were generated by using the shock size distribution.
The light-green curves and circles with error bars depict measurements of the cosmic 
background derived from {\em Akari}, COBE/DIRBE and COBE/FIRAS. Red squares
represent an estimation of the CIB excess after removing the contribution 
of the IGL obtained by stacking analysis (see text for details).
}
\label{fig:background2}
\end{figure*}

We have also analysed the impact of increasing $D$. Following \citet{cen2014},
we considered cases with $D=0.06$ and $D=0.4$. Results are displayed
in Fig.~\ref{fig:background2}. As expected from Equ.~\ref{eq:L_nu}, 
the radiation intensity increases almost proportional to $D$.  In particular,
in the extreme case of $D=0.4$, for both standard and shock grain-size distributions,
the predicted background emission from high redshifts could reach 1\% of
the measured flux. 
Besides, for extremely high $D$, dust emission from first galaxies
could reach $\sim 10$\% of the source-subtracted EBL in the case of 
the standard size distribution for dust grains and almost $\sim 100$\%
if assuming the shock size distribution.  
In the lower panels of Fig.~\ref{fig:background2}, we can see that 
the EBL also increases almost proportionally to $Z_{\rm g}$ (Equ.~\ref{eq:L_nu});
thus, $D$ and $Z_{\rm g}$ are degenerate quantities in our model.
If $Z_{\rm g}$ could be determined by other means, 
the strong dependence of the emission on
$D$ suggests that EBL measurements at sub-mm wavelengths
might provide important constraints on the amount of dust in the early Universe.

Finally, we explored the impact of increasing the star formation efficiency 
to an extreme value of $\eta \sim 0.1$.
As a higher ${\eta}$ would result in an enhanced ISM metallicity, we also increased 
$Z_{\rm g}$ to $0.05$, which is consistent with the closed-box model approximation
$Z = - y \ln ( 1 - {\eta})$, with $y$ being the stellar yield. Although this 
approximation could not represent the real behaviour of these systems, 
it provides an upper limit for the predicted EBL.
The obtained EBL spectrum is shown as a thick dashed black curve in Fig.~\ref{fig:background2},
bottom panels. The model EBL is below the observed source-subtracted EBL by
a factor of $\sim 10$ in the case of the standard size distribution for dust grains
(left panel).  On the other hand, for the shock size distribution (right panel), the model EBL
obtained under the aforementioned extreme assumptions reaches the observed EBL excess.

In summary, an increase of $D$, $Z_{\rm g}$ and $\eta$ 
from their fiducial values leads to a more important contribution of first galaxies
 to the FIR/sub-mm-EBL.  
Variations in $D$ or $Z_{\rm g}$ produces similar
changes in the EBL over our whole wavelength range while variations in 
$\eta$ affects mainly the spectrum at ${\lambda}_{\rm obs} < 1000 \ {\mu}{\rm m}$.
The source-subtracted EBL levels are only reached if extremely high values
are assumed for these parameters.
\citet{mitchellwynne2015} found that the total intensity of 
radiation emerging from galaxies at $z>6$ is 
$\log \nu I_{\nu } = -0.32 \pm 0.12$ at $1.6 \mu {\rm m}$, 
with a dominant contribution from $z > 8$ sources. These
authors constrained the cosmic luminosity density at $z>8$ to be
$\log {\rho}_{\rm UV} = 27.4^{+0.2}_{-1.2} \ {\rm erg \ s^{-1} \ Hz^{-1}
\ Mpc^{-3}}$, which is encouragingly similar to previous results
derived by \citet{kashlinsky2007} by using {\em Spitzer} data.
These findings imply that a substantial fraction of the 
UV radiation at the epoch of reionization is associated with sources well below
the sensitivity of current surveys.  
Here, we found that the contribution 
of these galaxies to the FIR-EBL would be even lower because of the
low dust mass density in the early Universe. 
Hence, individual point-source
detections of primordial dust sources are probably far beyond the capability of current instruments and upcoming facilities
(Fig. \ref{fig:spectra_z10}). 
Our results are consistent with \citet{fujimoto2015} who
analysed faint $1.2$ mm ALMA sources with a flux density of 
$\sim 0.01-1$ mJy, concluding that the total integrated $1.2$ mm flux
corresponds to $104^{+27}_{-30}$ \% of the EBL, measured by the COBE satellite.
These authors concluded that the dominant $1.2$ mm EBL contributors might be sources with $\gtrsim 0.01$ mJy.

\section{The NIR-EBL}

It is well-known that roughly 
half of the stellar light ever emitted during the history of the
Universe has been reprocessed by dust \citep{hauser2001}. 
As the objective of the present work is 
the study of the FIR/sub-mm-EBL signatures associated with the first galaxies, our model is
specially designed to describe the physics of dust emission.
However, this model does not include the physical prescriptions required to predict the NIR-EBL, such as
nebular emission or the reprocessing of Lyman-$\alpha$ photons by the intergalactic medium
\cite[e.g.][]{santos2002}.  Nevertheless, for the sake of comparison, we performed a rough estimation 
of the plausible contribution of our model galaxies to the NIR-EBL, assuming the same star formation rate densities as assumed in our FIR-EBL analysis.  For this purpose,
we implemented the simple formulation given in
\citet[][section 3.2]{greif2006}, but adapted to our own model parameters.  
The model links the ionizing photon production to the stellar mass inside halos:

\begin{equation}
\label{eq:I_NIR}
I_{\rm NIR} = 
\frac{h c}{4 \pi m_{\rm H}} {\eta}_{\rm ion} 
\int_{z_{\rm min}}^{z_{\rm max}} {\Psi}_* (z) 
\left| \frac{{\rm d}t}{{\rm d}z} \right| \ {\rm d}z,
\end{equation}

where ${\eta}_{\rm ion}$ is the number of ionizing photons emitted per stellar
baryon and ${\Psi}_* (z)$ is the star formation rate at a given $z$.
We assumed ${\eta}_{\rm ion} = 4 \times 10^3$, appropriate for Pop~I/Pop~II stars
\citep[see table~1 in][]{greif2006}.  For simplicity, we did not consider feedback
effects when modelling the star formation rate. Thus,

\begin{equation}
\label{eq:Psi}
{\Psi}_* (z) = {\rho}_{\rm m} \ \frac{{\Omega}_{\rm b}}{{\Omega}_{\rm M}} \
\eta \ \left| \frac{{\rm d}F_{\rm col}(z)}{{\rm d}z} \right| \ 
\left| \frac{{\rm d}z}{{\rm d}t} \right|,
\end{equation}

where ${\rho}_{\rm m}$ is the total mass density of the background universe and
$F_{\rm col}(z)$ represents the collapsed fraction of mass available for
star formation:

\begin{equation}
\label{eq:F_col}
F_{\rm col}(z) = \frac{1}{{\rho}_{\rm m}} \int_{M_{\rm min}}^{M_{\rm max}}
M_{\rm vir} \ n_{\rm ST} (M_{\rm vir}, z) \ {\rm d}M_{\rm vir}
\end{equation}
Note that Equ. \ref{eq:Psi} provides an upper limit for ${\Psi}_*$.

Using Equ. \ref{eq:I_NIR}-\ref{eq:F_col}, we can obtain an estimate of the contribution
of our model galaxies to the NIR-EBL, which can be compared to the FIR/sub-mm-EBL predicted
by using Equ. \ref{eq:Inu}.  For our model parameters, 
we obtain $I_{\rm NIR} \approx 3.4 \times 10^{-3} \ {\rm nW \ m^{-2} \ sr^{-1}}$.  This
value is a factor of $\sim 10$ higher than our prediction for the FIR/sub-mm-EBL based on 
the standard size distribution for dust grains.  However, the predicted $I_{\rm NIR}$
reaches a similar order of magnitude to our FIR/sub-mm-EBL when assuming a shock
size distribution.
It is worth mentioning that the simple model of \citet{greif2006} combined with our model
parameters is not able to reproduce the NIR-EBL excess of 
$\gtrsim 1\ {\rm nW \ m^{-2} \ sr^{-1}}$, 
inferred from the {\it Spitzer}/IRAC data \citep[e.g.][]{kashlinsky2005b}. 
Even assuming a maximum star
formation efficiency ${\eta} = 1$, the NIR-EBL associated with our galaxy population
remains below $\sim 0.5 \ {\rm nW \ m^{-2} \ sr^{-1}}$.

\begin{figure*}
\begin{center}
\resizebox{7.5cm}{!}{\includegraphics{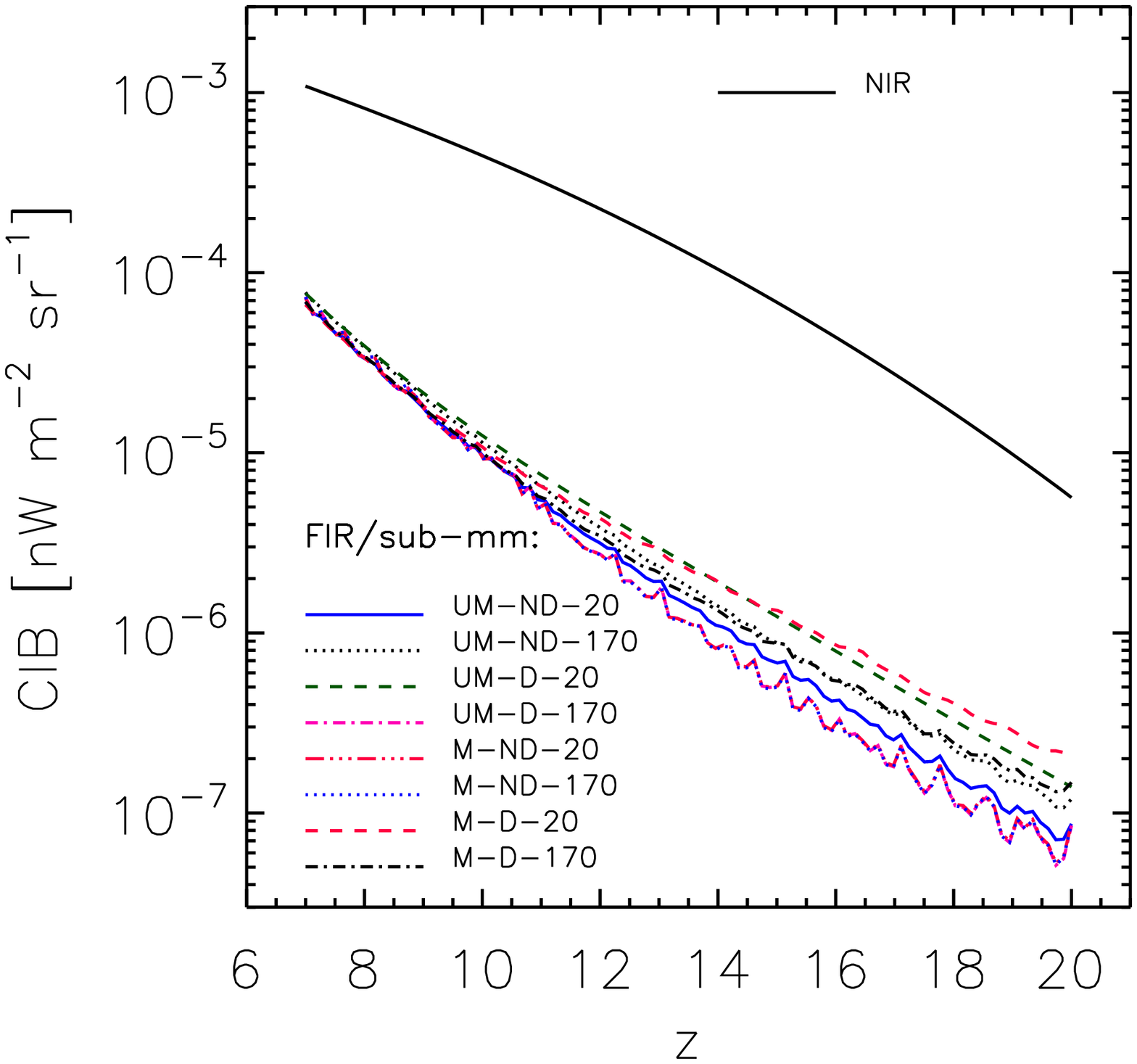}}
\resizebox{7.5cm}{!}{\includegraphics{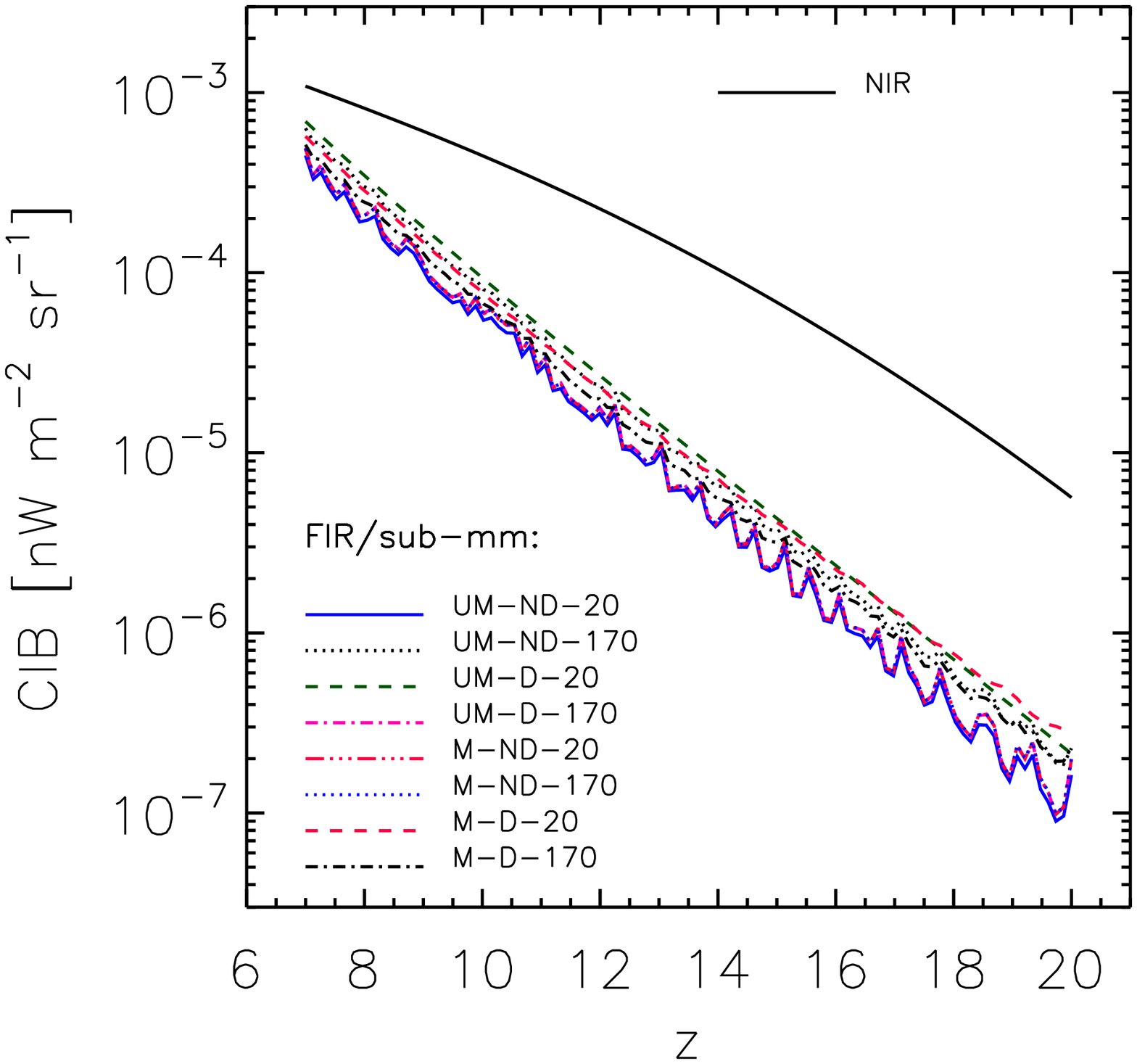}}\\
\end{center}
\caption[Background]
{ 
Contribution to the FIR/sub-mm-EBL of galaxy populations located at different $z$.  
Left and right panels show results derived from the standard and shock size distribution 
for dust grains, respectively. Findings for different dust chemical compositions
are indicated with different line-styles, as indicated in the figure.
The black solid lines depict the NIR-EBL obtained for our model galaxies by implementing
the prescription given in \citet{greif2006}; see text for details.
}
\label{fig:FIR_NIR}
\end{figure*}

In Fig. \ref{fig:FIR_NIR}, we compare the build-up of the CIB at NIR and FIR/sub-mm wavelengths 
associated with our model galaxies as a function of $z$. To obtain the distribution with $z$ of the FIR/sub-mm-EBL, 
we replaced the integral over $z$ in Equ. \ref{eq:Inu} with an integral over ${\nu}_{\rm obs}$ for
the interval ${\lambda}_{\rm obs}= 100 - 1000 \ {\mu}{\rm m}$.
We can see that, if assuming a standard size distribution for dust grains, the FIR/sub-mm-EBL
is a factor $\gtrsim 10$ below the NIR-EBL at all analysed $z$. On the other hand, 
for the shock size distribution, the ratio between the FIR/sub-mm and NIR EBL increases
from $\sim 0.03$ at $z\sim 20$ to $0.5$ at $z\sim 7$. 
As mentioned, a higher dust-to-metal ratio $D$ or a higher ISM metallicity $Z_{\rm g}$
would lead to an increase of the FIR/sub-mm-EBL and, consequently, of the ratio
between the FIR/sub-mm and NIR EBL.

Finally, according to a more recent and detailed analysis of \citet{helgason2016}, 
the contribution of galaxies at $z > 8$
to the NIR\--EBL could reach $\sim 0.01\--0.05 {\rm \ nW \ m^{-2} \ sr^{-1}}$.
If we compare these results with our findings regarding the FIR\--EBL, the ratio
between the FIR and NIR EBL at early times gives $\sim 0.002\--0.01$ and $\sim 0.02-0.1$ for the standard 
and shock size distribution, respectively. Evidently, these values are significantly lower than the fraction of
dust-reprocessed light inferred for lower $z$ ($\sim 50$\%).

\section{Summary and conclusions}
\label{sec:conclusions}

We have analysed the contribution of high-redshift galactic systems to the
FIR/sub\--mm EBL by implementing an idealized analytical model for dust emission
from the first galaxies. 
This model was then combined with the Sheth-Tormen halo mass function to estimate
the observable cosmic background produced by those sources.
We have considered different
dust chemical compositions and grain size distributions.

Our main results can be summarized as follows:

\begin{itemize}
\item For typical primeval dwarf\--size galaxies
($z \sim 10$, $M_{\rm vir} \sim 10^8 M_{\sun}$), the peak of dust emission
occurs at ${\lambda}_{\rm obs} \sim 500 {\mu} {\rm m}$.
A standard size distribution for dust grains generates a maximum observed flux of $\sim 10^{-3} {\rm nJy}$
while, in the case of a shock size distribution, observed fluxes are a factor $\lesssim 10$ higher.
According to our model, the observed flux originating from a typical first galaxy would be below the instrumental capabilities
of current and upcoming instruments.
By extrapolating our models to higher masses, we find that a system of at least
$10^{14} M_{\sun}$ is required to achieve detectability with the sensitivities of Herschel
and ALMA. Also, increasing our conservative dust-to-metal ratio ($D = 5 \times 10^{-3}$) or
ISM metallicity ($Z_{\rm g} = 5 \times 10^{-3} Z_{\sun}$)
would boost the resulting flux proportionally. A higher star formation efficiency
than our fiducial value ($\eta = 0.01$) would also yield a higher flux, 
specially at ${\lambda}_{\rm obs} < 1000 {\mu}{\rm m}$. 
\item By integrating the cumulative dust emission from sources with
$M_{\rm vir} \ge 10^7 M_{\sun}$ located at $z=7 \-- 20$, we obtained the
FIR/sub\--mm EBL emerging from the first galaxies.
The model EBL peaks at ${\lambda}_{\rm obs} \sim 500 \mu {\rm m}$ with an
intensity of $\sim 10^{-4}$ and $\sim 10^{-3} {\rm nW \ m^{-2} \ sr^{-1}}$ 
for the standard and shock size distribution of dust grains, respectively.
Accordingly, dust emission from these systems would not contribute significantly
to the measured cosmic FIR/sub-mm background, which exhibits values about $\sim 3 \-- 4$ 
orders of magnitude higher.
The low fluxes at FIR/sub\--mm wavelengths obtained for primeval galaxies
are consistent with different observational constraints indicating that
more massive and luminous galaxies at lower $z$ might be the main contributors to the
measured EBL \citep[e.g.][]{viero2013, leiton2015}.  
\item The detailed chemical composition of dust has a non significant impact on the obtained FIR/sub\--mm EBL, 
at least for the dust models studied here.
In particular, in the Rayleigh-Jeans regime, the slope of the model EBL spectrum
approaches the observed one. Thus, sources contributing to the observed EBL emission
would exhibit similar dust opacity curves as those used here.
\item By analysing the luminosity density as a function of the mass and redshift of model
galaxies, we found that the bulk of the FIR/sub\--mm EBL is associated with dwarf galaxies
($M_{\rm vir} \lesssim 10^{10} M_{\sun}$) at $z= 7\--20$, while more massive galaxies at
$z \sim 7$ tend to contribute mostly to the background at shorter wavelengths.
\item  Because of the low ISM densities of model galaxies, our results
seem to be robust against moderate variations in gas temperature or composition.
Also, we obtained no significant variations in our results when changing from the isothermal
to the NFW and Burkert gas density profiles. 
However, the model EBL exhibits a strong correlation with the dust-to-metal ratio $D$,
increasing almost proportionally with it.
Hence, measurements of the FIR/sub\--mm EBL could provide important
constraints on the amount of dust at early epochs.
\item By implementing a simple scheme, we made a rough estimation of the plausible contribution
of our model galaxies to the NIR-EBL. We obtained 
$I_{\rm NIR} \approx 3 \times 10^{-3} \ {\rm nW \ m^{-2} \ sr^{-1}}$, which agrees with the values
predicted for the FIR/sub-mm-EBL when using a shock size distribution for dust grains.
In the case of the standard size distribution, the predicted FIR/sub-mm flux is below $I_{\rm NIR}$
by a factor of $\sim 10$.
If we compare results from the more sophisticated NIR model by \citet{helgason2016} with our findings regarding 
the FIR\--EBL, the ratio between the FIR and NIR EBL at early times gives 
$\sim 0.002\--0.01$ and $\sim 0.02-0.1$ for the standard
and shock size distribution, respectively. 
\end{itemize}

Finally, measurements of the EBL are affected by the presence of significant
foregrounds at FIR/sub-mm wavelengths: zodiacal light and Galactic cirrus at $\lambda < 300 \mu {\rm m}$
and the cosmic microwave background at longer wavelengths.
Thus, absolute flux measurements of the FIR-EBL constitutes a very challenging task.
Although the low contribution to the net flux might prevent such absolute measurements of primordial
dust emission, it might still be possible to obtain information about it 
by exploring the EBL angular power spectrum
\citep{kashlinsky1996a, kashlinsky1996b, kashlinsky2000, arendt2010, mitchellwynne2015}, which we will investigate in a separate paper.  
According to our model, these signatures 
should be detected at $\sim 500 \mu {\rm m}$, close to the peak of dust emission from primeval sources. The overall hope is to open up complementary
windows into the different phases of the high-redshift star formation process,
by matching state-of-the-art theoretical predictions to the powerful
array of next-generation observational facilities.

\section*{Acknowledgements}
We thank the referee of this paper for useful suggestions and comments that
improved the manuscript.
We thank Alexander Ji for providing tabulated dust opacities for the different
dust models used here.
This work makes use of the Yggdrasil code \citep{zackrisson2011}, which adopts
Starburst99 SSP models, based on Padova-AGB tracks \citep{leitherer1999, vazquez2005}
for Population~II stars. VB would like to thank the IAFE in Buenos Aires for
its hospitality during the early stages of this work, and acknowledges
support from NSF grant AST-1413501.
MEDR is grateful to PICT-2015-3125 of ANPCyT (Argentina) and also to 
Mar\'{\i}a Sanz and Guadalupe Lucia for their help and support.

\bibliographystyle{mn2efix.bst}

\bibliography{references}

\begin{thebibliography}{105}
\expandafter\ifx\csname natexlab\endcsname\relax\def\natexlab#1{#1}\fi

\bibitem[{{Abel}, {Bryan} \& {Norman}(2002){Abel}, {Bryan}, \&
  {Norman}}]{abel2002}
{Abel} T., {Bryan} G.~L., {Norman} M.~L., 2002, Science, 295, 93

\bibitem[{{Allende Prieto}, {Lambert} \& {Asplund}(2001){Allende Prieto},
  {Lambert}, \& {Asplund}}]{allendeprieto2001}
{Allende Prieto} C., {Lambert} D.~L., {Asplund} M., 2001, \apjl, 556, L63

\bibitem[{{Arendt} {et~al}\mbox{.}(2010){Arendt}, {Kashlinsky}, {Moseley}, \&
  {Mather}}]{arendt2010}
{Arendt} R.~G., {Kashlinsky} A., {Moseley} S.~H., {Mather} J., 2010, \apjs,
  186, 10

\bibitem[{{Asplund}, {Grevesse} \& {Sauval}(2005){Asplund}, {Grevesse}, \&
  {Sauval}}]{asplund2005}
{Asplund} M., {Grevesse} N., {Sauval} A.~J., 2005, in Astronomical Society of
  the Pacific Conference Series, Vol. 336, Cosmic Abundances as Records of
  Stellar Evolution and Nucleosynthesis, {Barnes} III T.~G., {Bash} F.~N.,
  eds., p.~25

\bibitem[{{Asplund} {et~al}\mbox{.}(2009){Asplund}, {Grevesse}, {Sauval}, \&
  {Scott}}]{asplund2009}
{Asplund} M., {Grevesse} N., {Sauval} A.~J., {Scott} P., 2009, \araa, 47, 481

\bibitem[{{Barkana} \& {Loeb}(2001)}]{barkana2001}
{Barkana} R., {Loeb} A., 2001, \physrep, 349, 125

\bibitem[{{Beichman} \& {Helou}(1991)}]{beichman1991}
{Beichman} C.~A., {Helou} G., 1991, \apjl, 370, L1

\bibitem[{{Berta} {et~al}\mbox{.}(2010){Berta}, {Magnelli}, {Lutz}, {Altieri},
  {Aussel}, {Andreani}, {Bauer}, {Bongiovanni}, {Cava}, {Cepa}, {Cimatti},
  {Daddi}, {Dominguez}, {Elbaz}, {Feuchtgruber}, {F{\"o}rster Schreiber},
  {Genzel}, {Gruppioni}, {Katterloher}, {Magdis}, {Maiolino}, {Nordon},
  {P{\'e}rez Garc{\'{\i}}a}, {Poglitsch}, {Popesso}, {Pozzi}, {Riguccini},
  {Rodighiero}, {Saintonge}, {Santini}, {Sanchez-Portal}, {Shao}, {Sturm},
  {Tacconi}, {Valtchanov}, {Wetzstein}, \& {Wieprecht}}]{berta2010}
{Berta} S. {et~al.}, 2010, \aap, 518, L30

\bibitem[{{B{\'e}thermin} {et~al}\mbox{.}(2010){B{\'e}thermin}, {Dole},
  {Cousin}, \& {Bavouzet}}]{bethermin2010}
{B{\'e}thermin} M., {Dole} H., {Cousin} M., {Bavouzet} N., 2010, \aap, 516, A43

\bibitem[{{B{\'e}thermin} {et~al}\mbox{.}(2012){B{\'e}thermin}, {Le Floc'h},
  {Ilbert}, {Conley}, {Lagache}, {Amblard}, {Arumugam}, {Aussel}, {Berta},
  {Bock}, {Boselli}, {Buat}, {Casey}, {Castro-Rodr{\'{\i}}guez}, {Cava},
  {Clements}, {Cooray}, {Dowell}, {Eales}, {Farrah}, {Franceschini}, {Glenn},
  {Griffin}, {Hatziminaoglou}, {Heinis}, {Ibar}, {Ivison}, {Kartaltepe},
  {Levenson}, {Magdis}, {Marchetti}, {Marsden}, {Nguyen}, {O'Halloran},
  {Oliver}, {Omont}, {Page}, {Panuzzo}, {Papageorgiou}, {Pearson},
  {P{\'e}rez-Fournon}, {Pohlen}, {Rigopoulou}, {Roseboom}, {Rowan-Robinson},
  {Salvato}, {Schulz}, {Scott}, {Seymour}, {Shupe}, {Smith}, {Symeonidis},
  {Trichas}, {Tugwell}, {Vaccari}, {Valtchanov}, {Vieira}, {Viero}, {Wang},
  {Xu}, \& {Zemcov}}]{bethermin2012}
{B{\'e}thermin} M. {et~al.}, 2012, \aap, 542, A58

\bibitem[{{Bianchi} \& {Schneider}(2007)}]{bianchi2007}
{Bianchi} S., {Schneider} R., 2007, \mnras, 378, 973

\bibitem[{{Binney} \& {Tremaine}(2008)}]{binney2008}
{Binney} J., {Tremaine} S., 2008, {Galactic Dynamics: Second Edition}.
  Princeton University Press

\bibitem[{{Bromm}(2013{\natexlab{a}})}]{bromm2013}
{Bromm} V., 2013{\natexlab{a}}, Reports on Progress in Physics, 76, 112901

\bibitem[{{Bromm}(2013{\natexlab{b}})}]{bromm2011b}
{Bromm} V., 2013{\natexlab{b}}, Asociacion Argentina de Astronomia La Plata
  Argentina Book Series, 4, 3

\bibitem[{{Bromm}, {Coppi} \& {Larson}(2002){Bromm}, {Coppi}, \&
  {Larson}}]{bromm2002}
{Bromm} V., {Coppi} P.~S., {Larson} R.~B., 2002, \apj, 564, 23

\bibitem[{{Bromm} \& {Larson}(2004)}]{bromm2004}
{Bromm} V., {Larson} R.~B., 2004, \araa, 42, 79

\bibitem[{{Bromm} \& {Loeb}(2003)}]{bromm2003}
{Bromm} V., {Loeb} A., 2003, \nat, 425, 812

\bibitem[{{Bromm} \& {Yoshida}(2011)}]{bromm2011}
{Bromm} V., {Yoshida} N., 2011, \araa, 49, 373

\bibitem[{{Bromm} {et~al}\mbox{.}(2009){Bromm}, {Yoshida}, {Hernquist}, \&
  {McKee}}]{bromm2009}
{Bromm} V., {Yoshida} N., {Hernquist} L., {McKee} C.~F., 2009, \nat, 459, 49

\bibitem[{{Burkert}(1995)}]{burkert1995}
{Burkert} A., 1995, \apjl, 447, L25

\bibitem[{{Cai} {et~al}\mbox{.}(2013){Cai}, {Lapi}, {Xia}, {De Zotti},
  {Negrello}, {Gruppioni}, {Rigby}, {Castex}, {Delabrouille}, \&
  {Danese}}]{cai2013}
{Cai} Z.-Y. {et~al.}, 2013, \apj, 768, 21

\bibitem[{{Carniani} {et~al}\mbox{.}(2015){Carniani}, {Maiolino}, {De Zotti},
  {Negrello}, {Marconi}, {Bothwell}, {Capak}, {Carilli}, {Castellano},
  {Cristiani}, {Ferrara}, {Fontana}, {Gallerani}, {Jones}, {Ohta}, {Ota},
  {Pentericci}, {Santini}, {Sheth}, {Vallini}, {Vanzella}, {Wagg}, \&
  {Williams}}]{carniani2015}
{Carniani} S. {et~al.}, 2015, \aap, 584, A78

\bibitem[{{Carr}, {Bond} \& {Arnett}(1984){Carr}, {Bond}, \&
  {Arnett}}]{carr1984}
{Carr} B.~J., {Bond} J.~R., {Arnett} W.~D., 1984, \apj, 277, 445

\bibitem[{{Casey}, {Narayanan} \& {Cooray}(2014){Casey}, {Narayanan}, \&
  {Cooray}}]{casey2014}
{Casey} C.~M., {Narayanan} D., {Cooray} A., 2014, \physrep, 541, 45

\bibitem[{{Cen} \& {Kimm}(2014)}]{cen2014}
{Cen} R., {Kimm} T., 2014, \apj, 782, 32

\bibitem[{{Cherchneff} \& {Dwek}(2010)}]{cherchneff2010}
{Cherchneff} I., {Dwek} E., 2010, \apj, 713, 1

\bibitem[{{Chiaki} {et~al}\mbox{.}(2015){Chiaki}, {Marassi}, {Nozawa},
  {Yoshida}, {Schneider}, {Omukai}, {Limongi}, \& {Chieffi}}]{chiaki2015}
{Chiaki} G., {Marassi} S., {Nozawa} T., {Yoshida} N., {Schneider} R., {Omukai}
  K., {Limongi} M., {Chieffi} A., 2015, \mnras, 446, 2659

\bibitem[{{Condon}(1974)}]{condon1974}
{Condon} J.~J., 1974, \apj, 188, 279

\bibitem[{{Cooray} {et~al}\mbox{.}(2004){Cooray}, {Bock}, {Keatin}, {Lange}, \&
  {Matsumoto}}]{cooray2004}
{Cooray} A., {Bock} J.~J., {Keatin} B., {Lange} A.~E., {Matsumoto} T., 2004,
  \apj, 606, 611

\bibitem[{{Cooray} \& {Yoshida}(2004)}]{cooray2004b}
{Cooray} A., {Yoshida} N., 2004, \mnras, 351, L71

\bibitem[{{Coppin} {et~al}\mbox{.}(2006){Coppin}, {Chapin}, {Mortier}, {Scott},
  {Borys}, {Dunlop}, {Halpern}, {Hughes}, {Pope}, {Scott}, {Serjeant}, {Wagg},
  {Alexander}, {Almaini}, {Aretxaga}, {Babbedge}, {Best}, {Blain}, {Chapman},
  {Clements}, {Crawford}, {Dunne}, {Eales}, {Edge}, {Farrah}, {Gazta{\~n}aga},
  {Gear}, {Granato}, {Greve}, {Fox}, {Ivison}, {Jarvis}, {Jenness}, {Lacey},
  {Lepage}, {Mann}, {Marsden}, {Martinez-Sansigre}, {Oliver}, {Page},
  {Peacock}, {Pearson}, {Percival}, {Priddey}, {Rawlings}, {Rowan-Robinson},
  {Savage}, {Seigar}, {Sekiguchi}, {Silva}, {Simpson}, {Smail}, {Stevens},
  {Takagi}, {Vaccari}, {van Kampen}, \& {Willott}}]{Coppin2006}
{Coppin} K. {et~al.}, 2006, \mnras, 372, 1621

\bibitem[{{Couchman} \& {Rees}(1986)}]{couchman1986}
{Couchman} H.~M.~P., {Rees} M.~J., 1986, \mnras, 221, 53

\bibitem[{{Dole} {et~al}\mbox{.}(2006){Dole}, {Lagache}, {Puget}, {Caputi},
  {Fern{\'a}ndez-Conde}, {Le Floc'h}, {Papovich}, {P{\'e}rez-Gonz{\'a}lez},
  {Rieke}, \& {Blaylock}}]{dole2006}
{Dole} H. {et~al.}, 2006, \aap, 451, 417

\bibitem[{{Dopcke} {et~al}\mbox{.}(2013){Dopcke}, {Glover}, {Clark}, \&
  {Klessen}}]{dopcke2013}
{Dopcke} G., {Glover} S.~C.~O., {Clark} P.~C., {Klessen} R.~S., 2013, \apj,
  766, 103

\bibitem[{{Draine}(2011)}]{draine2011}
{Draine} B.~T., 2011, {Physics of the Interstellar and Intergalactic Medium}

\bibitem[{{Draine} \& {Li}(2001)}]{draine2001}
{Draine} B.~T., {Li} A., 2001, \apj, 551, 807

\bibitem[{{Dwek} {et~al}\mbox{.}(1998){Dwek}, {Arendt}, {Hauser}, {Fixsen},
  {Kelsall}, {Leisawitz}, {Pei}, {Wright}, {Mather}, {Moseley}, {Odegard},
  {Shafer}, {Silverberg}, \& {Weiland}}]{dwek1998}
{Dwek} E. {et~al.}, 1998, \apj, 508, 106

\bibitem[{{Dwek} \& {Krennrich}(2013)}]{dwek2012}
{Dwek} E., {Krennrich} F., 2013, Astroparticle Physics, 43, 112

\bibitem[{{Eales} {et~al}\mbox{.}(2000){Eales}, {Lilly}, {Webb}, {Dunne},
  {Gear}, {Clements}, \& {Yun}}]{eales1999}
{Eales} S., {Lilly} S., {Webb} T., {Dunne} L., {Gear} W., {Clements} D., {Yun}
  M., 2000, \aj, 120, 2244

\bibitem[{{Ferland} {et~al}\mbox{.}(1998){Ferland}, {Korista}, {Verner},
  {Ferguson}, {Kingdon}, \& {Verner}}]{ferland1998}
{Ferland} G.~J., {Korista} K.~T., {Verner} D.~A., {Ferguson} J.~W., {Kingdon}
  J.~B., {Verner} E.~M., 1998, \pasp, 110, 761

\bibitem[{{Finkbeiner}, {Davis} \& {Schlegel}(2000){Finkbeiner}, {Davis}, \&
  {Schlegel}}]{finkbeiner2000}
{Finkbeiner} D.~P., {Davis} M., {Schlegel} D.~J., 2000, \apj, 544, 81

\bibitem[{{Fixsen} {et~al}\mbox{.}(1998){Fixsen}, {Dwek}, {Mather}, {Bennett},
  \& {Shafer}}]{fixsen1998}
{Fixsen} D.~J., {Dwek} E., {Mather} J.~C., {Bennett} C.~L., {Shafer} R.~A.,
  1998, \apj, 508, 123

\bibitem[{{Frebel}, {Johnson} \& {Bromm}(2007){Frebel}, {Johnson}, \&
  {Bromm}}]{frebel2007}
{Frebel} A., {Johnson} J.~L., {Bromm} V., 2007, \mnras, 380, L40

\bibitem[{{Fujimoto} {et~al}\mbox{.}(2016){Fujimoto}, {Ouchi}, {Ono},
  {Shibuya}, {Ishigaki}, {Nagai}, \& {Momose}}]{fujimoto2015}
{Fujimoto} S., {Ouchi} M., {Ono} Y., {Shibuya} T., {Ishigaki} M., {Nagai} H.,
  {Momose} R., 2016, \apjs, 222, 1

\bibitem[{{Furlanetto} \& {Loeb}(2003)}]{furlanetto2003}
{Furlanetto} S.~R., {Loeb} A., 2003, \apj, 588, 18

\bibitem[{{Gall}, {Hjorth} \& {Andersen}(2011){Gall}, {Hjorth}, \&
  {Andersen}}]{gall2011}
{Gall} C., {Hjorth} J., {Andersen} A.~C., 2011, \aapr, 19, 43

\bibitem[{{Greif} \& {Bromm}(2006)}]{greif2006}
{Greif} T.~H., {Bromm} V., 2006, \mnras, 373, 128

\bibitem[{{Greif} {et~al}\mbox{.}(2011){Greif}, {Springel}, {White}, {Glover},
  {Clark}, {Smith}, {Klessen}, \& {Bromm}}]{greif2011}
{Greif} T.~H., {Springel} V., {White} S.~D.~M., {Glover} S.~C.~O., {Clark}
  P.~C., {Smith} R.~J., {Klessen} R.~S., {Bromm} V., 2011, \apj, 737, 75

\bibitem[{{Haiman}, {Thoul} \& {Loeb}(1996){Haiman}, {Thoul}, \&
  {Loeb}}]{haiman1996}
{Haiman} Z., {Thoul} A.~A., {Loeb} A., 1996, \apj, 464, 523

\bibitem[{{Hatsukade} {et~al}\mbox{.}(2013){Hatsukade}, {Ohta}, {Seko}, {Yabe},
  \& {Akiyama}}]{hatsukade2013}
{Hatsukade} B., {Ohta} K., {Seko} A., {Yabe} K., {Akiyama} M., 2013, \apjl,
  769, L27

\bibitem[{{Hauser} {et~al}\mbox{.}(1998){Hauser}, {Arendt}, {Kelsall}, {Dwek},
  {Odegard}, {Weiland}, {Freudenreich}, {Reach}, {Silverberg}, {Moseley},
  {Pei}, {Lubin}, {Mather}, {Shafer}, {Smoot}, {Weiss}, {Wilkinson}, \&
  {Wright}}]{hauser1998}
{Hauser} M.~G. {et~al.}, 1998, \apj, 508, 25

\bibitem[{{Hauser} \& {Dwek}(2001)}]{hauser2001}
{Hauser} M.~G., {Dwek} E., 2001, \araa, 39, 249

\bibitem[{{Helgason} {et~al}\mbox{.}(2016){Helgason}, {Ricotti}, {Kashlinsky},
  \& {Bromm}}]{helgason2016}
{Helgason} K., {Ricotti} M., {Kashlinsky} A., {Bromm} V., 2016, \mnras, 455,
  282

\bibitem[{{Hollenbach} \& {McKee}(1979)}]{hollenbach1979}
{Hollenbach} D., {McKee} C.~F., 1979, \apjs, 41, 555

\bibitem[{{Ji}, {Frebel} \& {Bromm}(2014){Ji}, {Frebel}, \& {Bromm}}]{ji2014}
{Ji} A.~P., {Frebel} A., {Bromm} V., 2014, \apj, 782, 95

\bibitem[{{Karlsson}, {Bromm} \& {Bland-Hawthorn}(2013){Karlsson}, {Bromm}, \&
  {Bland-Hawthorn}}]{karlsson2013}
{Karlsson} T., {Bromm} V., {Bland-Hawthorn} J., 2013, Reviews of Modern
  Physics, 85, 809

\bibitem[{{Kashlinsky}(2005)}]{kashlinsky2005}
{Kashlinsky} A., 2005, \physrep, 409, 361

\bibitem[{{Kashlinsky} {et~al}\mbox{.}(2004){Kashlinsky}, {Arendt}, {Gardner},
  {Mather}, \& {Moseley}}]{kashlinsky2004}
{Kashlinsky} A., {Arendt} R., {Gardner} J.~P., {Mather} J.~C., {Moseley} S.~H.,
  2004, \apj, 608, 1

\bibitem[{{Kashlinsky} {et~al}\mbox{.}(2005){Kashlinsky}, {Arendt}, {Mather},
  \& {Moseley}}]{kashlinsky2005b}
{Kashlinsky} A., {Arendt} R.~G., {Mather} J., {Moseley} S.~H., 2005, \nat, 438,
  45

\bibitem[{{Kashlinsky} {et~al}\mbox{.}(2007){Kashlinsky}, {Arendt}, {Mather},
  \& {Moseley}}]{kashlinsky2007}
{Kashlinsky} A., {Arendt} R.~G., {Mather} J., {Moseley} S.~H., 2007, \apjl,
  654, L1

\bibitem[{{Kashlinsky} {et~al}\mbox{.}(2015){Kashlinsky}, {Mather}, {Helgason},
  {Arendt}, {Bromm}, \& {Moseley}}]{kashlinsky2015}
{Kashlinsky} A., {Mather} J.~C., {Helgason} K., {Arendt} R.~G., {Bromm} V.,
  {Moseley} S.~H., 2015, \apj, 804, 99

\bibitem[{{Kashlinsky}, {Mather} \& {Odenwald}(1996){Kashlinsky}, {Mather}, \&
  {Odenwald}}]{kashlinsky1996a}
{Kashlinsky} A., {Mather} J.~C., {Odenwald} S., 1996, \apjl, 473, L9

\bibitem[{{Kashlinsky} {et~al}\mbox{.}(1996){Kashlinsky}, {Mather}, {Odenwald},
  \& {Hauser}}]{kashlinsky1996b}
{Kashlinsky} A., {Mather} J.~C., {Odenwald} S., {Hauser} M.~G., 1996, \apj,
  470, 681

\bibitem[{{Kashlinsky} \& {Odenwald}(2000)}]{kashlinsky2000}
{Kashlinsky} A., {Odenwald} S., 2000, \apj, 528, 74

\bibitem[{{Kashlinsky} {et~al}\mbox{.}(2002){Kashlinsky}, {Odenwald}, {Mather},
  {Skrutskie}, \& {Cutri}}]{kashlinsky2002}
{Kashlinsky} A., {Odenwald} S., {Mather} J., {Skrutskie} M.~F., {Cutri} R.~M.,
  2002, \apjl, 579, L53

\bibitem[{{Kaufman}(1976)}]{kaufman1976}
{Kaufman} M., 1976, \apss, 40

\bibitem[{{Kormendy} {et~al}\mbox{.}(2009){Kormendy}, {Fisher}, {Cornell}, \&
  {Bender}}]{kormendy2009}
{Kormendy} J., {Fisher} D.~B., {Cornell} M.~E., {Bender} R., 2009, \apjs, 182,
  216

\bibitem[{{Larson}(1969)}]{larson1969}
{Larson} R.~B., 1969, \mnras, 145, 405

\bibitem[{{Leitherer} {et~al}\mbox{.}(1999){Leitherer}, {Schaerer}, {Goldader},
  {Delgado}, {Robert}, {Kune}, {de Mello}, {Devost}, \&
  {Heckman}}]{leitherer1999}
{Leitherer} C. {et~al.}, 1999, \apjs, 123, 3

\bibitem[{{Leiton} {et~al}\mbox{.}(2015){Leiton}, {Elbaz}, {Okumura}, {Hwang},
  {Magdis}, {Magnelli}, {Valtchanov}, {Dickinson}, {B{\'e}thermin},
  {Schreiber}, {Charmandaris}, {Dole}, {Juneau}, {Le Borgne}, {Pannella},
  {Pope}, \& {Popesso}}]{leiton2015}
{Leiton} R. {et~al.}, 2015, \aap, 579, A93

\bibitem[{{Loeb}(2010)}]{loeb2010}
{Loeb} A., 2010, {How Did the First Stars and Galaxies Form? Princeton Univ.
  Press, Princeton }

\bibitem[{{Loeb} \& {Haiman}(1997)}]{loeb1997}
{Loeb} A., {Haiman} Z., 1997, \apj, 490, 571

\bibitem[{{Low} \& {Tucker}(1968)}]{low1968}
{Low} F.~J., {Tucker} W.~H., 1968, Physical Review Letters, 21, 1538

\bibitem[{{Magliocchetti}, {Salvaterra} \& {Ferrara}(2003){Magliocchetti},
  {Salvaterra}, \& {Ferrara}}]{magliocchetti2003}
{Magliocchetti} M., {Salvaterra} R., {Ferrara} A., 2003, \mnras, 342, L25

\bibitem[{{Marsden} {et~al}\mbox{.}(2009){Marsden}, {Ade}, {Bock}, {Chapin},
  {Devlin}, {Dicker}, {Griffin}, {Gundersen}, {Halpern}, {Hargrave}, {Hughes},
  {Klein}, {Mauskopf}, {Magnelli}, {Moncelsi}, {Netterfield}, {Ngo}, {Olmi},
  {Pascale}, {Patanchon}, {Rex}, {Scott}, {Semisch}, {Thomas}, {Truch},
  {Tucker}, {Tucker}, {Viero}, \& {Wiebe}}]{marsden2009}
{Marsden} G. {et~al.}, 2009, \apj, 707, 1729

\bibitem[{{Matsuura} {et~al}\mbox{.}(2011){Matsuura}, {Shirahata}, {Kawada},
  {Takeuchi}, {Burgarella}, {Clements}, {Jeong}, {Hanami}, {Khan}, {Matsuhara},
  {Nakagawa}, {Oyabu}, {Pearson}, {Pollo}, {Serjeant}, {Takagi}, \&
  {White}}]{matsuura2011}
{Matsuura} S. {et~al.}, 2011, \apj, 737, 2

\bibitem[{{Mayer} \& {Duschl}(2005)}]{mayer2005}
{Mayer} M., {Duschl} W.~J., 2005, \mnras, 358, 614

\bibitem[{{Mitchell-Wynne} {et~al}\mbox{.}(2015){Mitchell-Wynne}, {Cooray},
  {Gong}, {Ashby}, {Dolch}, {Ferguson}, {Finkelstein}, {Grogin}, {Kocevski},
  {Koekemoer}, {Primack}, \& {Smidt}}]{mitchellwynne2015}
{Mitchell-Wynne} K. {et~al.}, 2015, Nature Communications, 6, 7945

\bibitem[{{Navarro}, {Frenk} \& {White}(1997){Navarro}, {Frenk}, \&
  {White}}]{navarro1997}
{Navarro} J.~F., {Frenk} C.~S., {White} S.~D.~M., 1997, \apj, 490, 493

\bibitem[{{Nozawa} \& {Kozasa}(2013)}]{nozawa2013}
{Nozawa} T., {Kozasa} T., 2013, \apj, 776, 24

\bibitem[{{Omukai}(2000)}]{omukai2000}
{Omukai} K., 2000, \apj, 534, 809

\bibitem[{{Omukai}, {Hosokawa} \& {Yoshida}(2010){Omukai}, {Hosokawa}, \&
  {Yoshida}}]{omukai2010}
{Omukai} K., {Hosokawa} T., {Yoshida} N., 2010, \apj, 722, 1793

\bibitem[{{Ono} {et~al}\mbox{.}(2014){Ono}, {Ouchi}, {Kurono}, \&
  {Momose}}]{ono2014}
{Ono} Y., {Ouchi} M., {Kurono} Y., {Momose} R., 2014, \apj, 795, 5

\bibitem[{{Planck Collaboration} {et~al}\mbox{.}(2014){Planck Collaboration},
  {Ade}, {Aghanim}, {Armitage-Caplan}, {Arnaud}, {Ashdown}, {Atrio-Barandela},
  {Aumont}, {Baccigalupi}, {Banday}, \& et~al.}]{planck2014}
{Planck Collaboration} {et~al.}, 2014, \aap, 571, A16

\bibitem[{{Pollack} {et~al}\mbox{.}(1994){Pollack}, {Hollenbach}, {Beckwith},
  {Simonelli}, {Roush}, \& {Fong}}]{pollack1994}
{Pollack} J.~B., {Hollenbach} D., {Beckwith} S., {Simonelli} D.~P., {Roush} T.,
  {Fong} W., 1994, \apj, 421, 615

\bibitem[{{Puget} {et~al}\mbox{.}(1996){Puget}, {Abergel}, {Bernard},
  {Boulanger}, {Burton}, {Desert}, \& {Hartmann}}]{puget1996}
{Puget} J.-L., {Abergel} A., {Bernard} J.-P., {Boulanger} F., {Burton} W.~B.,
  {Desert} F.-X., {Hartmann} D., 1996, \aap, 308, L5

\bibitem[{{Safranek-Shrader}, {Bromm} \&
  {Milosavljevi{\'c}}(2010){Safranek-Shrader}, {Bromm}, \&
  {Milosavljevi{\'c}}}]{safranek2010}
{Safranek-Shrader} C., {Bromm} V., {Milosavljevi{\'c}} M., 2010, \apj, 723,
  1568

\bibitem[{{Safranek-Shrader}, {Milosavljevi{\'c}} \&
  {Bromm}(2014){Safranek-Shrader}, {Milosavljevi{\'c}}, \&
  {Bromm}}]{safranek2014}
{Safranek-Shrader} C., {Milosavljevi{\'c}} M., {Bromm} V., 2014, \mnras, 438,
  1669

\bibitem[{{Salvaterra} \& {Ferrara}(2003)}]{salvaterra2003}
{Salvaterra} R., {Ferrara} A., 2003, \mnras, 339, 973

\bibitem[{{Salvaterra} \& {Ferrara}(2006)}]{salvaterra2006}
{Salvaterra} R., {Ferrara} A., 2006, \mnras, 367, L11

\bibitem[{{Santos}, {Bromm} \& {Kamionkowski}(2002){Santos}, {Bromm}, \&
  {Kamionkowski}}]{santos2002}
{Santos} M.~R., {Bromm} V., {Kamionkowski} M., 2002, \mnras, 336, 1082

\bibitem[{{Schlegel}, {Finkbeiner} \& {Davis}(1998){Schlegel}, {Finkbeiner}, \&
  {Davis}}]{schlegel1998}
{Schlegel} D.~J., {Finkbeiner} D.~P., {Davis} M., 1998, \apj, 500, 525

\bibitem[{{Schneider}, {Ferrara} \& {Salvaterra}(2004){Schneider}, {Ferrara},
  \& {Salvaterra}}]{schneider2004}
{Schneider} R., {Ferrara} A., {Salvaterra} R., 2004, \mnras, 351, 1379

\bibitem[{{Schneider} {et~al}\mbox{.}(2006){Schneider}, {Omukai}, {Inoue}, \&
  {Ferrara}}]{schneider2006_methods}
{Schneider} R., {Omukai} K., {Inoue} A.~K., {Ferrara} A., 2006, \mnras, 369,
  1437

\bibitem[{{Sheth}, {Mo} \& {Tormen}(2001){Sheth}, {Mo}, \&
  {Tormen}}]{sheth2001}
{Sheth} R.~K., {Mo} H.~J., {Tormen} G., 2001, \mnras, 323, 1

\bibitem[{{Smith} {et~al}\mbox{.}(2015){Smith}, {Safranek-Shrader}, {Bromm}, \&
  {Milosavljevi{\'c}}}]{smith2015}
{Smith} A., {Safranek-Shrader} C., {Bromm} V., {Milosavljevi{\'c}} M., 2015,
  \mnras, 449, 4336

\bibitem[{{Stacy}, {Greif} \& {Bromm}(2010){Stacy}, {Greif}, \&
  {Bromm}}]{stacy2010}
{Stacy} A., {Greif} T.~H., {Bromm} V., 2010, \mnras, 403, 45

\bibitem[{{Tegmark} {et~al}\mbox{.}(1997){Tegmark}, {Silk}, {Rees},
  {Blanchard}, {Abel}, \& {Palla}}]{tegmark1997}
{Tegmark} M., {Silk} J., {Rees} M.~J., {Blanchard} A., {Abel} T., {Palla} F.,
  1997, \apj, 474, 1

\bibitem[{{V{\'a}zquez} \& {Leitherer}(2005)}]{vazquez2005}
{V{\'a}zquez} G.~A., {Leitherer} C., 2005, \apj, 621, 695

\bibitem[{{Viero} {et~al}\mbox{.}(2013){Viero}, {Moncelsi}, {Quadri},
  {Arumugam}, {Assef}, {B{\'e}thermin}, {Bock}, {Bridge}, {Casey}, {Conley},
  {Cooray}, {Farrah}, {Glenn}, {Heinis}, {Ibar}, {Ikarashi}, {Ivison}, {Kohno},
  {Marsden}, {Oliver}, {Roseboom}, {Schulz}, {Scott}, {Serra}, {Vaccari},
  {Vieira}, {Wang}, {Wardlow}, {Wilson}, {Yun}, \& {Zemcov}}]{viero2013}
{Viero} M.~P. {et~al.}, 2013, \apj, 779, 32

\bibitem[{{Wei{\ss}} {et~al}\mbox{.}(2009){Wei{\ss}}, {Kov{\'a}cs}, {Coppin},
  {Greve}, {Walter}, {Smail}, {Dunlop}, {Knudsen}, {Alexander}, {Bertoldi},
  {Brandt}, {Chapman}, {Cox}, {Dannerbauer}, {De Breuck}, {Gawiser}, {Ivison},
  {Lutz}, {Menten}, {Koekemoer}, {Kreysa}, {Kurczynski}, {Rix}, {Schinnerer},
  \& {van der Werf}}]{weiss2009}
{Wei{\ss}} A. {et~al.}, 2009, \apj, 707, 1201

\bibitem[{{Wilson} {et~al}\mbox{.}(2008){Wilson}, {Austermann}, {Perera},
  {Scott}, {Ade}, {Bock}, {Glenn}, {Golwala}, {Kim}, {Kang}, {Lydon},
  {Mauskopf}, {Predmore}, {Roberts}, {Souccar}, \& {Yun}}]{Wilson2008}
{Wilson} G.~W. {et~al.}, 2008, \mnras, 386, 807

\bibitem[{{Yoshida} {et~al}\mbox{.}(2003){Yoshida}, {Abel}, {Hernquist}, \&
  {Sugiyama}}]{yoshida2003}
{Yoshida} N., {Abel} T., {Hernquist} L., {Sugiyama} N., 2003, \apj, 592, 645

\bibitem[{{Yoshida}, {Bromm} \& {Hernquist}(2004){Yoshida}, {Bromm}, \&
  {Hernquist}}]{yoshida2004}
{Yoshida} N., {Bromm} V., {Hernquist} L., 2004, \apj, 605, 579

\bibitem[{{Zackrisson} {et~al}\mbox{.}(2011){Zackrisson}, {Rydberg},
  {Schaerer}, {{\"O}stlin}, \& {Tuli}}]{zackrisson2011}
{Zackrisson} E., {Rydberg} C.-E., {Schaerer} D., {{\"O}stlin} G., {Tuli} M.,
  2011, \apj, 740, 13

\end{thebibliography}

\end{document}